\documentstyle[prd,epsf,preprint,aps]{revtex}

\begin{document}
\draft

\title{Constituent quark model for nuclear stopping in high energy nuclear collisions}
\author{T. K. Choi, M. Maruyama and  F. Takagi}
\address{Department of physics, Tohoku University, Sendai, 980-77, Japan}

\date{\today}
\maketitle

\begin{abstract}
  We study the nuclear stopping in high energy nuclear collisions using the constituent quark model. 
It is assumed that wounded nucleons with different number of interacted quarks  hadronize in different ways.
The probabilities of having such wounded nucleons are evaluated for proton-proton, proton-nucleus and nucleus-nucleus collisions.
After examining our model in proton-proton and proton-nucleus collisions and fixing the hadronization functions, it is extended to nucleus-nucleus collisions.
It is used to calculate the rapidity distribution and the rapidity shift of final state protons in nucleus-nucleus collisions.
The computed results are in good agreement with the experimental data on $^{32}\mbox{S} +\ ^{32}\mbox{S}$ at $E_{lab} = 200$ AGeV and  $^{208}\mbox{Pb} +\ ^{208}\mbox{Pb}$ at $E_{lab} = 160$ AGeV.
Theoretical predictions are also given for proton rapidity distribution in $^{197}\mbox{Au} +\ ^{197}\mbox{Au}$ at $\sqrt{s} = 200$ AGeV (BNL-RHIC).
We predict that the nearly baryon free region will appear in the midrapidity region and the rapidity shift is $\langle \Delta y \rangle = 2.22$.
\end{abstract}

\pacs{25.75, 12.39, 25.40.E, 13.85}

\section{Introduction}
  Whether the incident nucleons are stopped in or passed through the target nucleus is a fundamental and important concern in high energy heavy ion collisions. 
These two different situations emerge according to collision energy and atomic mass of nucleus.
They suggest different relevant dynamics, the dynamics of shock formation and Landau hydrodynamics \cite{lan53} in case of stopping regime or the Bjorken longitudinal expansion and inside-outside cascade dynamics \cite{bjo83} in case of baryon free regime.
The distinction between the two situations is related to how the incident nucleons slow down by multiple collision with nucleons of the other nucleus and how the collision energy are deposited for particle production.
The relevant measure, the nuclear stopping power provides an estimation of the energy density achieved in collisions.
Experimental data indicate so far a baryon rich regime at midrapidity for heavy nucleus collision such as gold on gold at AGS energies \cite{barr92} and sulphur on sulphur \cite{bac94} or lead on lead \cite{esu95} at SPS, where the achievement of initial condition for QGP formation is still controversial.
The clear baryon free regime may be realized at RHIC energy region. 
\par
For the description of high energy nucleus collision dynamics, there are many theoretical models which differ in their assumptions as to how the particles share the incident energies, where the sources of particle production are and how the produced particles hadronize, while having almost the same picture for a multiple collision process of constituents such as the Glauber model \cite{gla59}.
A detailed comparison of these models can be found in Ref. \cite{wer93} by Werner and Ref. \cite{won94} by Wong.
In particular, for the proton rapidity density distribution which is related to how high baryon densities may be attained in reactions, these model predictions show notable difference due to their different assumptions.
For example, as discussed by Gyulassy \cite{gyu95}, the RQMD \cite{sor89} predicts a much higher degree of baryon stopping than VENUS \cite{wer93} and HIJING \cite{xnw91} which predict a concavity at midrapidity for a reaction such as lead on lead collision at SPS energies. 
In some approaches a nucleon-nucleon model is directly extended to nucleus-nucleus collisions without examining proton-nucleus collisions.
Our opinion is that model for nuclear collisions should first be examined for $NA$ collisions and then extended to $AB$ collisions.
\par
The notion of constituent quarks as the units of collision has been shown to be very useful to describe not only hadron-hadron collisions but also hadron-nucleus($hA$) collisions at high energies.
In particular, the successful results has been obtained in application to the projectile fragmentation region of $hA$ collisions, since the first application to $hA$ collisions by Anisovich {\it{et al}} \cite{ani78} and, later on, further developed by several authors \cite{nik79,dar80,bia79,tak81,tak83}.
\par

In this paper, We first formulate the constituent quark model(CQM) relevant to nucleon-nucleon, nucleon-nucleus and nucleus-nucleus collision dynamics at high energy by considering quarks, the constituents of hadrons, as the unit of collision instead of hadrons.
When the projectile nucleon is incident on the target in nucleon-nucleon collision, the collided nucleon has the probability of becoming three types of wounded nucleon with one, two or all the three interacted quarks. 
In case of nucleon-nucleus collision, the incident nucleon after collision has these probabilities by multiple collision with individual nucleons within the target nucleus. 
In nucleus-nucleus collision, each collided nucleon from the projectile nucleus becomes one of the three types of wounded nucleons. 
This distinction of the three types of wounded nucleons is a crucial difference from multiple-collision models constructed at the nucleon level.
The three probabilities of quark absorption can be calculated from the given nuclear density and the total inelastic quark-quark cross section $\sigma_{inel}^{qq}$.
\par

The three probabilities are used to estimate the degree of nuclear stopping or the rapidity distribution in high energy nucleus-nucleus collisions. 
The rapidity density distribution or the momentum distribution in nuclear collision is expressed by three factors, the flavor factor, the quark interaction probability of the incident nucleon and the fragmentation function. 
The flavor factor is introduced to make a distinction whether a final state baryon is an observed proton, a neutron or a nucleon decayed from a hyperon.
The determination of fragmentation function $f_{i}(x)$ comes from fitting the data on proton-proton \cite{bre82} and proton-nucleus collision \cite{bai85}.
\par

This paper is organized as follows. In section \ref{sec:for}, we show the assumption of CQM and the quark interaction probabilities having three types of wounded nucleons with one, two or all the three interacted quarks in nucleon-nucleon collision at high energy. 
In section \ref{sec:pa}, the quark interaction probabilities in nucleon-nucleus collisions are given as a function of the mass number $A$ in nucleon-nucleus collision.
In section \ref{sec:aa}, the average numbers of three type wounded nucleons are calculated for $AA$ interaction. 
In section \ref{sec:x}, we compute the fractional momentum distribution of protons in $pA$ collision.
In section \ref{sec:y}, we calculate the rapidity distribution of leading protons in $AA$ interaction. 
Finally, section \ref{sec:cd} is devoted to conclusions and discussions. 

\newpage

\section{Quark interactions in nucleon-nucleon collision}
\label{sec:for}
We start giving the outline of the assumptions of the CQM pertinent to the description of nuclear stopping in high energy nuclear collisions.
The CQM is based on three fundamental assumptions which are related to the structure of hadrons, the interactions between constituents of the projectile and the target hadrons and the hadronization of quarks in participant nucleons \cite{ani78,nik79,dar80,bia79,tak81,tak83}.
We first assume that a hadron(meson or baryon) consists of two or three spatially separated constituent quarks. 
Secondly, in a hadron-hadron, hadron-nucleus or nucleus-nucleus collision, some quarks from the projectile are assumed to interact independently with some quarks from the target thus losing a considerable fraction of their initial momenta, while the quarks which escape from colliding in both the projectile and the target pass through retaining their initial momenta.
The third assumption claims that those quarks hadronize eventually via fragmentation and recombination mechanism.
\par

In order to calculate the total inelastic cross section for nucleon-nucleon collisions in terms of CQM, we need the probability of having a quark-quark inelastic collision when one quark in the projectile is at an impact parameter $\bbox{b}$ relative to another quark in the target which is given by
\begin{equation}
h(\bbox{b})= \sigma_{inel}^{qq}\delta^{(2)}(\bbox{b})
\label{eqn:d1}
\end{equation}
in the point particle approximation.
The integration over impact parameter gives the total inelastic cross section of quark-quark collisions;
\begin{equation}
\int h(\bbox{b})d\bbox{b}= \sigma_{inel}^{qq}.
\label{eqn:d2}
\end{equation}
\par

We consider the collision of a beam nucleon $N$ with a target nucleon $N'$.
To the probability of a quark-quark collision Eq.\ (\ref{eqn:d1}),
by multiplying the probability elements for finding a quark 
$\rho_{N}(\bbox{b}_{N})d\bbox{b}_{N}dz_{N}$
and
$\rho_{N'}(\bbox{b}_{N'})d\bbox{b}_{N'}dz_{N'}$
in the volume element
$d\bbox{b}_{N}dz_{N}$
and
$d\bbox{b}_{N'}dz_{N'}$,
and by integrating over the collision axis $z_{N}$ and $z_{N'}$,  
we obtain the probability that a particular quark in $N$ interact with a particular quark in $N'$ when $N$ and $N'$ are at an impact parameter
$\bbox{b}$
relative to each other,
\begin{equation}
W(\bbox{b})= 
\int 
d\bbox{b}_{N} 
d\bbox{b}_{N'}
\rho_{N}(\bbox{b}_{N})
\rho_{N'}(\bbox{b}_{N'})
h(\bbox{b}-\bbox{b}_{N}+\bbox{b}_{N'}),
\label{eqn:d3}
\end{equation}
where 
$\rho_{N}(\bbox{b}_{N})$
is the $z_{N}$-integrated normalized quark distribution in the nucleon;
 \begin{equation}
\rho_{N}(\bbox{b}_{N})=\int \rho_{N}(\bbox{b}_{N},z_{N})dz_{N}.
\label{eqn:d4}
\end{equation}
It is normalized as
\begin{equation}
\int \rho_{N}(\bbox{b}_{N})d\bbox{b}_{N}=1. 
\label{eqn:d5}
\end{equation}
\par

Using 
$W(\bbox{b})$ of (\ref{eqn:d3}), we can evaluate the total inelastic cross section and various probabilities.
When a projectile nucleon is at an impact parameter $\bbox{b}$ relative to a target nucleon, the probability of the occurrence of an inelastic event is
\begin{equation}
g(\bbox{b})=
1-\{ 1-W(\bbox{b}) \}^{9},
\label{eqn:d6}
\end{equation}
where the second term means the probability that all quarks in the projectile pass through the target nucleon without any inelastic collision.
Therefore, we obtain the total inelastic cross section $\sigma^{NN'}_{inel}$ for  $NN'$ collision
\begin{equation}
\sigma^{NN'}_{inel}=\int
d\bbox{b}
 g(\bbox{b}).
\label{eqn:d7}
\end{equation}
The probability $g(\bbox{b})$ of Eq.\ (\ref{eqn:d6}) can also be expressed as a sum of $g^{(i)}(\bbox{b})$ which is the probability that $i$ quarks in the projectile nucleon $N$ collide with any quarks of the target nucleon $N'$;
\begin{equation}
g(\bbox{b})=\sum^{3}_{i=1}g^{(i)}(\bbox{b}),
\label{eqn:d9}
\end{equation}
where
\begin{equation}
g^{(i)}(\bbox{b})=
{{3}\choose{i}}
\left[1-\{1-W(\bbox{b}) \}^{3}\right]^{i}
\left[1-W(\bbox{b})\right]^{3(3-i)}.
\label{eqn:d8}
\end{equation}
Here ${{3}\choose{i}}$ is the combinatorial factor and the term $1-\{1-W(\bbox{b}) \}^{3}$ implies the probability that a particular quark out of the projectile nucleon interact at least once with the quarks in the target nucleon.
Integrating over the impact parameter $\bbox{b}$ and dividing by the total inelastic cross section, we obtain three integrated probabilities that $i$ quarks out of three are absorbed in nucleon-nucleon collision;
\begin{equation}
P_{NN'}^{(i)}=
\frac{1}{\sigma^{NN'}_{inel}}
\int d\bbox{b} 
g^{(i)}(\bbox{b}),
\label{eqn:d10}
\end{equation}
which is illustrated in Fig.\ \ref{fig:figppa}.
\par

Given the quark density distribution $\rho_{N}(\bbox{b})$ and the inelastic quark-quark cross section $\sigma_{inel}^{qq}$, we can calculate the total inelastic proton-proton cross section $\sigma_{inel}^{pp}$ and the probability of quark absorption $P_{pp}^{(i)}$ in proton-proton collision.
The quark distribution is assumed to be Gaussian for simplicity,
\begin{equation}
\rho_{N}(\bbox{b})=
\frac{1}{2\pi\beta^{2}}\exp(-\frac{\bbox{b}^{2}}{2\beta^{2}}), 
\label{eqn:d11}
\end{equation}
where the parameter $\beta$ is related to the root-mean-square radius of the proton $r_{rms}^{p}$,
\begin{equation}
\beta^{2}=\frac{1}{3}(r_{rms}^{p})^{2}.
\label{eqn:d12}
\end{equation}
Electron scattering data \cite{sim80} gives $r_{rms}^{p}=0.862$ fm. 
For the inelastic quark-quark cross section as the input, we take $\sigma_{inel}^{qq}=4.32$ mb in order to reproduce the total inelastic cross section of $pp$ collision $\sigma_{inel}^{pp}=30$ mb which is observed for center-of-mass energy 3 GeV $\lesssim$ $\sqrt{s}$ $\lesssim$ 100 GeV and $\sigma_{inel}^{qq}=6.64$ mb to give $\sigma_{in}^{pp}=41$ mb for $\sqrt{s} = $ 200 GeV \cite{hik92}.
The numerical values of the probability of quark interaction in $pp$ collision for $\sigma_{inel}^{qq} = 4.32$mb are shown in Table\ \ref{tab:pacp}.
It should be noticed that $P^{(2)}_{pp}$ is considerably large implying violation of the additive quark approximation \cite{bia82}.

\newpage

\section{Quark interactions in nucleon-nucleus collision}
\label{sec:pa}
When an incident nucleon collide with the target nucleus, the projectile nucleon can interact with many nucleons in the nucleus. 
We use the probabilities of quark absorption in $NN$ collision to obtain those in $NA$ interactions.
When a nucleon is incident at an impact parameter 
$\bbox{b}$
relative to the nucleus $A$, the probability for the nucleon to collide with a particular nucleon in the target nucleus is given by
\begin{equation}
V_{A}(\bbox{b})= 
\int 
d\bbox{b}_{A} 
\rho_{A}(\bbox{b}_{A})
g(\bbox{b}-\bbox{b}_{A}),
\label{eqn:f1}
\end{equation}
where 
$\rho_{A}(\bbox{b}_{A})$ is the $z$-integrated nucleon density distribution of nucleus $A$,
$d\bbox{b}_{A} \rho_{A}(\bbox{b}_{A})$
is the $z$-integrated probability element of finding the nucleon and 
$g(\bbox{b}-\bbox{b}_{A})$
is the probability of inelastic $NN$ interactions at impact parameter 
$\bbox{b}-\bbox{b}_{A}$
given by Eq.\ (\ref{eqn:d6}).
\par

The total inelastic cross section for $NA$ collisions is given by
\begin{equation}
\sigma^{NA}_{inel}=\int
d\bbox{b}
\left[1-\{ 1-V_{A}(\bbox{b}) \}^{A}\right].
\label{eqn:f2}
\end{equation} 
As $g(\bbox{b}-\bbox{b}_{A})$ in Eq.\ (\ref{eqn:f1}) is given by (\ref{eqn:d9}),  $V_{A}(\bbox{b})$ is decomposed into a sum of three terms;
\begin{equation}
V_{A}(\bbox{b})=
\sum_{i=1}^{3}V_{A}^{(i)}(\bbox{b})
\label{eqn:f4}
\end{equation}
where $V_{A}^{(i)}(\bbox{b})$ is the probability that $i$ quarks in the projectile nucleon interact with a nucleon in the target nucleus,
\begin{equation}
V_{A}^{(i)}(\bbox{b})= 
\int 
d\bbox{b}_{A} 
\rho_{A}(\bbox{b}_{A})
g^{(i)}(\bbox{b}-\bbox{b}_{A}).
\label{eqn:f3}
\end{equation}
Let us calculate the probabilities $P_{NA}^{(j)}$ for projectile nucleon having $j$ interacted quarks in $NA$ collisions.
We first expand $\sigma^{NA}_{inel}$ of Eq.\ (\ref{eqn:f2}) into a sum of the contributions from n $N-N$ collisions;
\begin{equation}
\sigma^{NA}_{inel}= \sum_{n=1}^{A}
\int d\bbox{b} 
{{A}\choose{n}}
\left[V_{A}(\bbox{b})\right]^{n}
\left[1-V_{A}(\bbox{b}) \right]^{A-n}.
\label{eqn:f31}
\end{equation}
By substituting Eq.\ (\ref{eqn:f4}) for (\ref{eqn:f31}) and expanding the latter, we obtain 
\begin{equation}
P_{NA}^{(j)}=
\frac{1}{\sigma^{NA}_{inel}}
\sum_{n=1}^{A}
{A\choose n}
\int d\bbox{b} 
U_{NA}^{(j)}(n;\bbox{b})
\{1-V_{A}(\bbox{b}) \}^{A-n},
\label{eqn:f5}
\end{equation}
where
$U_{NA}^{(j)}(n;\bbox{b})$ is 
the probability of having $j$ interacted quarks in $n$ $N-N$ collisions and is given by
\begin{mathletters}
\label{eqn:allf5ac}
\begin{eqnarray}
U_{NA}^{(1)}(n;\bbox{b})
&=& U_{(1)}^{(1)}(n;\bbox{b}); \\
U_{NA}^{(2)}(n;\bbox{b})
\label{eqn:f5a}
&=& U_{(1)}^{(2)}(n;\bbox{b})+U_{(2)}^{(2)}(n;\bbox{b})\\ \nonumber
&&\mbox{     }+\sum_{k=1}^{n-1}{n\choose k}
\left[\frac23U_{(1)}^{(1)}(k;\bbox{b})U_{(2)}^{(2)}(n-k;\bbox{b})
+\frac13U_{(1)}^{(2)}(k;\bbox{b})U_{(2)}^{(2)}(n-k;\bbox{b})\right];   \\
U_{NA}^{(3)}(n;\bbox{b})
\label{eqn:f5b}
&=& \left[V_{A}(\bbox{b})\right]^{n} 
-U_{NA}^{(1)}(n;\bbox{b})-U_{NA}^{(2)}(n;\bbox{b}),
\label{eqn:f5c}
\end{eqnarray}   
\end{mathletters}
where
$U_{(i)}^{(j)}(n;\bbox{b})$ is the probability of having $j$ interacted quarks in $n$ $N-N$collisions while having $i$ interacted quarks in each collision and is given by
\begin{mathletters}
\label{eqn:allf6ac}
\begin{eqnarray}
U_{(1)}^{(1)}(n;\bbox{b})&=&3\left(\frac{1}{3}V_{A}^{(1)}(\bbox{b})\right)^{n}; \\
\label{eqn:f6a}
U_{(1)}^{(2)}(n;\bbox{b})&=&3(2^{n}-2)\left(\frac{1}{3}V_{A}^{(1)}(\bbox{b})\right)^{n}; \\
\label{eqn:f6b}
U_{(2)}^{(2)}(n;\bbox{b})&=&3\left(\frac{1}{3}V_{A}^{(2)}(\bbox{b})\right)^{n}.
\label{eqn:f6c}
\end{eqnarray}
\end{mathletters}
In the Appendix \ref{sec:appa} we present the details of our calculation of $P_{NA}^{(j)}$.
\par

Now we need to fix the nucleon density distribution $\rho_{A}(\bbox{b})$ in order to carry out the numerical calculation.
We use the Woods-Saxon parameterization, 
\begin{equation}
\rho_{A}(\bbox{r})=\frac{\rho_{0}}{1+\exp{(\frac{\bbox{r}-R}{a}})}
\label{eqn:f7}
\end{equation}
for heavy nuclei ($ A \geq 16$) while
the Gaussian distribution
\begin{equation}
\rho_{A}(r)=\frac{1}{(2\pi\beta^{2})^{3/2}}\exp(-\frac{r^{2}}{2\beta^{2}})
\label{eqn:f8}
\end{equation}
for $^{9}$Be,
where both distributions are normalized as
\begin{equation}
\int d^{3}r \rho_{A}(r)=1.
\label{eqn:f9}
\end{equation}
The parameters for each distribution can be obtained from the data on elastic electron and hadron scattering on nuclei.
For the Woods-Saxon distribution, we use the parameter $R$ and $a$ in Ref.\ \cite{bor69}:
\begin{equation}
R=1.12A^{\frac13}-0.86A^{-\frac{1}{3}},\qquad  a=0.54 \mbox{fm}
\label{eqn:f10}
\end{equation}
for ($ A \geq 16$).
For the Gaussian distribution, we take \cite{den73}:
\begin{equation}
\beta^{2}=\frac{1}{3}(r_{rms}^{A})^{2},\qquad  r_{rms}^{Be}=2.3 \mbox{fm}
\label{eqn:f11}
\end{equation}
for $^{9}$Be.
The calculated values of the total inelastic cross section
 $\sigma_{inel}^{pA}$ 
and the probabilities 
$P_{pA}^{(1)}$, $P_{pA}^{(2)}$ and $ P_{pA}^{(3)}$ 
 are shown in Table \ref{tab:pacp} and the behavior of $P_{pA}^{(i)}$ as a function of $A$ is shown in Fig.\ \ref{fig:pai}.
The calculated cross sections are in good agreement with the experimental values \cite{bai85,car79}. 
The process of one quark interaction with the probability $P_{pA}^{(1)}$ is dominant for light nuclei.
However, even for Be, the probability of having two interacting quarks is not so small. 
All the three probabilities have comparable magnitudes for $A \gtrsim 60$. 

\newpage

\section{Quark interactions in nucleus-nucleus collisions}
\label{sec:aa}
In nucleus-nucleus collision, many nucleons of the projectile nucleus($A$) may collide with many nucleons in the target nucleus($B$)  by multiple collisions. 
After the full process of multiple collisions, the incident nucleons are divided into four types of nucleons having zero, one, two or all the three interacted quarks.
When the beam nucleus $A$ and the target nucleus $B$ are situated at an impact parameter
$\bbox{b}$
relative to each other, the probability for the occurrence of a collision between one nucleon in $A$ and one nucleon in $B$ is given by
\begin{equation}
V_{AB}(\bbox{b})= 
\int 
d\bbox{b}_{A} 
d\bbox{b}_{B}
\rho_{A}(\bbox{b}_{A})
\rho_{B}(\bbox{b}_{B})
g(\bbox{b}+\bbox{b}_{A}-\bbox{b}_{B}),
\label{eqn:e1}
\end{equation}
where
$g(\bbox{b})$
is given by Eq.\ (\ref{eqn:d6}).
\par

The total inelastic cross section for $AB$ collision is given by
\begin{equation}
\sigma^{AB}_{inel}=\int
d\bbox{b}
\left[1-\{ 1-V_{AB}(\bbox{b}) \}^{AB}\right].
\label{eqn:e2}
\end{equation}
In the same way as in Eqs.\ (\ref{eqn:f4}) and (\ref{eqn:f3}), the probability 
$V_{AB}(\bbox{b})$
 can be expressed as a sum of three probabilities of incident nucleons having $i$ interacted quarks
\begin{equation}
V_{AB}(\bbox{b})=\sum_{i=1}^{3}V_{AB}^{(i)}(\bbox{b})
\label{eqn:e31}
\end{equation}
where
\begin{equation}
V_{AB}^{(i)}(\bbox{b})= 
\int 
d\bbox{b}_{A} 
d\bbox{b}_{B} 
\rho_{A}(\bbox{b}_{A})
\rho_{B}(\bbox{b}_{B})
g^{(i)}(\bbox{b}+\bbox{b}_{A}-\bbox{b}_{B}),
\label{eqn:e3}
\end{equation}
with
$g^{(i)}(\bbox{b}+\bbox{b}_{A}-\bbox{b}_{B})$
being given by Eq.\ (\ref{eqn:d8}).
The expansion of total inelastic cross section Eq.\ (\ref{eqn:e2}) gives the probabilities of having $m$ wounded nucleons on the beam side
\begin{equation}
P_{AB}(m)=
\frac{1}{\sigma^{AB}_{inel}}
\int d\bbox{b} 
{{A}\choose{m}}
\left[1-\{1-V_{AB}(\bbox{b}) \}^{A}\right]^{m}
\left[ 1-V_{AB}(\bbox{b}) \right]^{B(A-m)},
\label{eqn:e4}
\end{equation} 
and the average number is
\begin{equation}
\langle m \rangle = \sum_{m=1}^{A}mP_{AB}(m).
\label{eqn:e5}
\end{equation}
The number $m$ is a sum of $m_{i}$, the number of nucleons having $i$ interacted quarks;
\begin{equation}
m=\sum_{i=1}^{3}m_{i}.
\label{eqn:e51}
\end{equation}
We shall evaluate the probability $P_{AB}(m;m_{1},m_{2},m_{3})$ of having such configuration of wounded nucleons.
It is convenient to introduce the probability $R_{AB}^{(j)}(\bbox{b})$ that one of incident nucleons has $j$ interacted quarks at a fixed impact parameter $\bbox{b}$.
It is obtained in a similar way as in $NA$ collision by expanding the factor $1-\{1-V_{AB}(\bbox{b}) \}^{A}$.
The result is
\begin{equation}
R_{AB}^{(j)}(\bbox{b})=
\sum_{n=1}^{B}
{{B}\choose{n}}
U_{AB}^{(j)}(n;\bbox{b})
\left[1-V_{AB}(\bbox{b}) \}\right]^{B-n}
\label{eqn:e6}
\end{equation} 
with
\begin{equation}
\sum_{j=1}^{3}R_{AB}^{(j)}(\bbox{b})=
1-\{1-V_{AB}(\bbox{b}) \}^{B}.
\nonumber
\end{equation} 
Here, 
$U_{AB}^{(j)}(n;\bbox{b})$
is obtained by substituting 
$V_{AB}^{(i)}(\bbox{b})$
in $AB$ interaction
instead of
$V_{A}^{(i)}(\bbox{b})$
in $NA$ interaction in Eqs.\ (\ref{eqn:allf5ac}) and (\ref{eqn:allf6ac}).
Using the polynomial expansion, we have 
\begin{eqnarray}
\lefteqn{P_{AB}(m;m_{1},m_{2},m_{3})}   \nonumber \\
& &=\frac{1}{\sigma^{AB}_{inel}}
\int d\bbox{b} 
{{A}\choose{m}}{{m}\choose{m_{1}}}{{m-m_{1}}\choose{m_{2}}}  \nonumber \\
&&\mbox{      }\times
\left[R_{AB}^{(1)}(\bbox{b})\right]^{m_{1}}
\left[R_{AB}^{(2)}(\bbox{b})\right]^{m_{2}}
\left[R_{AB}^{(3)}(\bbox{b})\right]^{m_{3}}
\left[\{ 1-V_{AB}(\bbox{b}) \}^{B}\right]^{A-m},
\label{eqn:e7}
\end{eqnarray}
where $m_{3}= m-m_{1}-m_{2}$.
The average number of $m_{j}$ is given by,
\begin{equation}
\langle m_{j} \rangle = \sum_{m=1}^{A}\sum_{m_{1}=0}^{m}\sum_{m_{2}=0}^{m-m_{1}}m_{j}P_{AB}(m;m_{1},m_{2},m_{3}).
\label{eqn:e8}
\end{equation}
\par

Using the nucleon density distribution of Eq.\ (\ref{eqn:f7}) with Eq.\ (\ref{eqn:f10}), the nucleus-nucleus inelastic cross section and the average number of wounded nucleons are calculated for $^{32}S$ + $^{32}S$ ($E_{lab}=200$ AGeV), $^{208}Pb$ + $^{208}Pb$ ($E_{lab}=160$ AGeV) and $^{197}Au$ + $^{197}Au$ ($\sqrt{s}=200$ AGeV) reactions as there are recent experimental data from CERN SPS for the first two reactions.
The third reaction is chosen to give a prediction for RHIC experiment.
Along with $\sigma^{AA}_{inel}$, the calculated values of 
 $\langle m \rangle $, $\langle m_{1} \rangle $, $\langle m_{2} \rangle $ and $\langle m_{3} \rangle $ 
for no-bias events are listed in Table \ref{tab:aacp}.
For $Au$ + $Au$ and $Pb$ + $Pb$ reactions, 
we have shown the values estimated from the empirical formula
$\sigma^{AB}_{inel}=\pi r_{0}^{2}(A^{\frac13}+B^{\frac13}-b)^{2}$
with parameters $r_{0}= 1.48$ fm and $b=1.32$ fm \cite{abd80}.
The calculated cross section $\sigma^{AA}_{inel}$ is larger than the experimental values for $S$ + $S$, while it is in good agreement with data for $Au$ + $Au$ and $Pb$ + $Pb$ reactions.
The discrepancy of the cross section in $S$ + $S$ does not affect on 
$P_{AB}(m;m_{1},m_{2},m_{3})$ and $\langle m_{j} \rangle $.
\par

The probability distributions $P_{AB}(m)$ of Eq.\ (\ref{eqn:e4}) and the corresponding average values $\langle m \rangle$ for 
$S$ + $S$ and $Pb$ + $Pb$ reactions for different triggers
are shown, respectively, in Fig.\ \ref{fig:pm} and Table \ref{tab:aacp} where the values of $\langle m_{j} \rangle$ are shown also.
Here, Veto and $E_{T}$ triggers in $S$ + $S$ reactions mean central triggers which correspond to 2\% and 11\% of total inelastic cross section, respectively \cite{bac94}.
Corresponding to these triggers, we introduce the impact parameter cutoff $b_{max}$ in our theoretical calculations so that the partial cross section
\begin{equation}
\int_{|\bbox{b}| < b_{max}}
d\bbox{b}
\sigma_{inel}^{AB}(\bbox{b})
\nonumber
\end{equation}  
becomes 2\% or 11\% of the total cross section.
The cutoff is found to be 
$b_{max}=1.18$ fm for the Veto trigger and $b_{max}=2.78$ fm for the $E_{T}$ trigger.
For both $Pb$ + $Pb$ and $Au$ + $Au$ reactions, we take $b_{max}=6.05$ fm and $b_{max}=5.86$ fm, respectively, corresponding to 15\% of $\sigma_{inel}^{AA}$.
On the other hand, the peripheral trigger in $S$ + $S$ reactions correspond to 65\% of $\sigma_{inel}^{AA}$, which lead to the small $|\bbox{b}|$ cutoff $b_{min}= 4.9$ fm.
\par

In Fig.\ \ref{fig:pm}(a), the distribution $P_{AB}(m)$ for no-bias event of $S$ + $S$ reactions shows the well-known horse-back shape.
It is obvious that the distribution at small $m$ is dominated by peripheral collision while the one at large $m$ is by the central ones.
The clear separation of two components in the $m$-distribution is the consequence from the strong correlation between $m$ and $\bbox{b}$.
As shown in Fig.\ \ref{fig:pm}(b), in $Pb$ + $Pb$ collisions, the distribution for no-bias events show prominent peaks in both the smallest and the largest $m$-region.
The largest $m$-region is of course dominated by the central collisions.
The most remarkable features of the corresponding average value $\langle m \rangle$ shown in Table \ref{tab:aacp} is that more than 94\% of the incident nucleons are wounded in the central collisions of both $Pb$ + $Pb$ and $Au$ + $Au$.
\par

Shown in Fig.\ \ref{fig:pmi} are $m$-dependence of fractions $\langle m_{i}(m) \rangle/m$ in no-bias event of $S$ + $S$ and $Pb$ + $Pb$ reactions.
Here $\langle m_{i}(m) \rangle/m$ is the average value of $m_{i}$ for a fixed $m$.
At small $m$, the wounded nucleon having one interacted quark $m_{1}$ dominate in both reactions. 
At larger $m$, the fractions $\langle m_{1}(m) \rangle/m$ and $\langle m_{2}(m) \rangle/m$ are comparable to each other in $S$ + $S$ reaction.
On the other hand, for $m \gtrsim 197$, the fraction $\langle m_{3}(m) \rangle/m$ becomes largest in $Pb$ + $Pb$ reactions.
For this region, $\langle m_{3} \rangle$ is largest in central $Pb$ + $Pb$ collision at $\sqrt{s} = 17.4$ AGeV as seen in Table \ref{tab:aacp}.
The trend is stronger in $Au$ + $Au$ collisions at RHIC energy because of larger $\sigma_{inel}^{qq}$.

\newpage

\section{Fractional momentum distributions of protons in proton-nucleus collisions}
\label{sec:x}
In this section, we study the inclusive spectra of protons in the projectile fragmentation region of proton-nucleus collisions as it gives a basic information on the nuclear stopping in high energy nuclear collisions.
According to CQM, both participant quarks and spectator quarks hadronize via fragmentation and recombination.
However, instead of giving a detailed description of such hadronization dynamics, we here introduce a phenomenological fragmentation function $f_{i}(x)$ ($i = 1, 2, 3$) which is fractional momentum distribution of protons coming from a wounded nucleon having $i$ interacted quarks.
Assuming the independent fragmentation of each type of wounded nucleons, we express the proton spectra in $pp$, $pA$ and $AB$ interactions as
\begin{mathletters}
\label{eqn:allx2x4}
\begin{eqnarray}
\left.\frac{dN}{dx}\right|_{pp\rightarrow pX}&=&
\sum_{i=1}^{3}\lambda_{i}P_{pp}^{(i)}f_{i}(x), 
\label{eqn:x2} \\
\left.\frac{dN}{dx}\right|_{pA\rightarrow pX}&=&
\sum_{i=1}^{3}\lambda_{i}P_{pA}^{(i)}f_{i}(x), 
\label{eqn:x3} \\
\left.\frac{dN}{dx}\right|_{AB\rightarrow pX}&=&
\sum_{i=1}^{3}\lambda_{i}^{(A)}\langle m_{i} \rangle f_{i}(x),
\label{eqn:x4}
\end{eqnarray}
\end{mathletters}
for $0 < x < 1$, where $dN/dx$ is the normalized single particle inclusive cross section $\sigma^{-1}_{inel} d\sigma/dx$ and $x$ is the Feynman scaling variable defined in c.m.s.
Moreover, $\lambda_{i}$ and $\lambda_{i}^{(A)}$ are the flavor factors which can be interpreted as the probabilities of finding a proton in the hadronization product from wounded nucleon having $i$ interacted quarks provided the effect of baryon-antibaryon pair production is negligible.
The probabilities $P_{pp}^{(i)}$ and $P_{pA}^{(i)}$ are given by Eqs.\ (\ref{eqn:d10}) and (\ref{eqn:f5}) while $\langle m_{i} \rangle$ are by Eq.\ (\ref{eqn:e8}) and 
all the $f_{i}(x)$ are normalized as
\begin{equation}
\int_{0}^{1}dxf_{i}(x)=1.
\label{eqn:x5}
\end{equation}
It is obvious that the r.h.s of Eq.\ (\ref{eqn:allx2x4}), being integrated over $x$ from $0$ to $1$, just gives the average multiplicity of the final state protons for each reactions in the beam fragmentation region.
\par

Three fragmentation functions can be determined if data on three different reactions are given.
In addition to $pp \rightarrow pX$ of Ref.\ \cite{bre82}, we use data on $pCu \rightarrow pX$ and $pAg \rightarrow pX$ provided by two experimental groups \cite{bai85,bar83} in order to cover the wide $x$ range.
As Ref.\ \cite{bar83} gives only the cross sections at fixed $p_{T}$'s, we assumed that they are proportional to the $p_{T}$-integrated cross section;
\begin{equation}
 x\frac{d\sigma}{dx} = \kappa \left.E\frac{d\sigma}{d\mbox{\boldmath $p$}}\right|_{\mbox{fixed $p_{T}$}}.
\label{eqn:x10}
\end{equation}
Actually the constant $\kappa$ can be determined by requiring that the data of Ref.\ \cite{bar83} at $p_{T} = 0.3$ GeV/c coincide with the data of Ref.\ \cite{bai85} in the range $0.3 < x < 0.6$ where there exist data points from both groups.
Such a procedure is possible because the $x$-dependence of a fixed $p_{T}$-cross section of \cite{bar83} is really similar to $xd\sigma/dx$ of \cite{bai85} as shown in Figs.\ \ref{fig:pax} (c) and (d).
The values of $\kappa$ are $\kappa = 1.30$ and 1.20 [$(GeV/c)^{2}$] for $A = Cu$ and $Ag$, respectively.
For information, the same procedure has been applied to $pp \rightarrow pX$ with the result that $\kappa = 1.18$ [$(GeV/c)^{2}$].
See Fig.\ \ref{fig:pax} (a).
It is remarkable that the values of $\kappa$ are almost the same, $i.e.$, independent of $A$ in the three cases.
\par

Applying data on three different reactions to Eqs.\ (\ref{eqn:x2}) and (\ref{eqn:x3}), one can determine the three fragmentation functions.
However, the immediate application gives too large uncertainties for $f_{2}(x)$ and $f_{3}(x)$ due to the propagation of the experimental errors.
We accordingly assume that $f_{3}(x) \sim 0$ for $x \gtrsim 1/3$ because a wounded nucleon having all the three interacted quarks loses most of incident momentum.
The fragmentation functions thus obtained are shown as points with error bars in Fig.\ \ref{fig:frg}, where they represent the values of the fragmentation functions including the flavor factors, $F_{i}(x) = \lambda_{i}f_{i}(x)$ at various $x$.
\par

Though the fragmentation functions have been determined at various $x$, we need $F_{i}(x)$ for all $x$ in order to calculate the nuclear stopping power in any reactions.
In this case a meaningful $\chi^{2}$ fit for whole $x$ region is impossible because of absence of experimental data at $x < 0.2$ and the large errors of data point at $0.2 \leq x \leq 0.3$.
Therefore, we assume appropriate functional forms and impose some physical conditions: $F_{1}(x)$ is assumed to be a linear function of $x$, $F_{2}(x)$ and $F_{3}(x)$ are gaussian and exponential, respectively, with conditions on the average fractional momenta $\langle x \rangle_{1} > \langle x \rangle_{2} > \langle x \rangle_{3}$ and another condition on the flavor factors $\lambda_{1} > \lambda_{2} > \lambda_{3}$.
Here, we do not use the $pp \rightarrow pX$ data point at $x > 0.9$ shown in Fig.\ \ref{fig:pax}(a) because they are dominated by the diffractive contribution.
The inequality for the average fractional momenta $\langle x \rangle_{i} = \int dxxf_{i}(x)$ means that a wounded nucleon with the more interaction loses the more momentum.
The probability of the flavor change in the incident nucleon is larger for the more interacted quarks giving the inequality for the flavor factors.
The resultant fragmentation functions are 
\begin{mathletters}
\label{eqn:allx11x13}
\begin{eqnarray}
F_{1}(x)
&=& 0.56+0.18x, 
\label{eqn:x11}\\
F_{2}(x)
&=& 1.4\{\exp(-3.54x^{2})-\exp(-3.54)\},
\label{eqn:x12} \\
F_{3}(x)
&=& 3.25\exp(-6.52x).
\label{eqn:x13}
\end{eqnarray}
\end{mathletters}
As shown in Fig.\ \ref{fig:frg}, $F_{1}(x)$ is nearly constant while both $F_{2}(x)$ and $F_{3}(x)$ decrease toward zero as $x$ increases.
The flavor factors are $\lambda_{1}=0.65, \lambda_{2}=0.61$ and $\lambda_{3}=0.50$ and the average fractional momenta are $\langle x \rangle_{1} = 0.52$, $\langle x \rangle_{2} = 0.28$ and $\langle x \rangle_{3} = 0.15$.
\par

In Fig.\ \ref{fig:pax}, the calculated proton spectra for $pA \rightarrow pX$ reactions ($A = p, Be, Cu, Ag, W$ and $U$) are compared with the experimental data \cite{bre82} \cite{bai85} \cite{bar83}.
Our theoretical proton spectra estimated from Eq.\ (\ref{eqn:allx2x4}) with above fragmentation functions Eq.\ (\ref{eqn:allx11x13}) and probabilities of quark absorption given in Table\ \ref{tab:pacp} reproduce not only the input data ($p, Cu, Ag$) but also the other data ($Be, W, U$).
This results suggest the validity of CQM.

\newpage

\section{Rapidity distribution of protons in AA collisions}
\label{sec:y}
The rapidity distributions of participant protons have recently been measured for $S+S$ collisions at $E_{lab} = 200$ AGeV by NA35 Collaboration \cite{bac94} and also for central $^{208}Pb$ + $^{208}Pb$ collision at $ E_{lab}= 160$ AGeV by NA44 Collaboration \cite{esu95}.
In our model the fractional momentum distribution of protons can be calculated from Eq. (\ref{eqn:x4}) with $f_{i}(x)$ given by Eq. (\ref{eqn:allx11x13}), $\langle m_{i}\rangle$ given in Table \ref{tab:aacp} and the flavor factors $\lambda_{i}^{(A)}$ given in Table \ref{tab:flaf}.
Calculation of the factors $\lambda_{i}^{(A)}$ is given in Appendix \ref{sec:appb}.
The rapidity distribution can be obtained from the fractional momentum distribution (\ref{eqn:x4}) using the relation
\begin{equation}
\frac{dN}{dy}=
 \sqrt{x^{2}+\frac{4{m_{T}}^{2}}{s}}\frac{dN}{dx},
\label{eqn:y1}
\end{equation}
where $m_{T}$ is the transverse mass of the particle; ${m_{T}}^{2}=m^{2}+ \mbox{\boldmath $p$}_{T}^{2}$.
\par

The results for $S + S$ collisions are shown in Fig.\ \ref{fig:ys}.
We have used the average values  $ p_{T}$ = 0.622, 0.595 and 0.45 GeV/c for central(veto and $E_{T}$ trigger) and peripheral collisions, respectively \cite{bac94}.
There is a good agreement between theoretical results and experimental data, particularly in the midrapidity region.
The rise of the calculated spectra at very small $y_{lab}$ is due to the behavior of $f_{1}(x)$ at $x \simeq 1$ which is rather ambiguous.
There is a trend that the proton yield is slightly overestimated for the central triggers while it is underestimated for the peripheral collision.
This may be due to a sharp cut of the impact parameter in our calculation.
Our model can also reproduce well the preliminary data on central $Pb + Pb$ collision \cite{esu95} as shown in Fig.\ \ref{fig:ypb}.
Considerable yield of protons in the central rapidity region( $y_{lab} \simeq 3$ ) is due to the dominance of $\langle m_{3}(m)/m \rangle$ at the largest m. See Fig.\ \ref{fig:pmi}(b). 
\par

In Fig.\ \ref{fig:yau}, the theoretical predictions is given for  $^{197}Au$ + $^{197}Au$ collisions at $\sqrt{s} = 200$ AGeV to be measured at BNL-RHIC.
It is remarkable that the midrapidity region becomes nearly baryon free in contrast to $Pb + Pb$ collisions at CERN-SPS(Fig.\ \ref{fig:ypb}).
\par

As a useful measure of nuclear stopping power, one usually uses the mean rapidity shift of the projectile proton from their original beam rapidity, $\langle\Delta y\rangle$ or the mean fractional momentum retained by the final state proton, $\langle x \rangle$ \cite{dat85}.
The larger the $\langle\Delta y\rangle$, the more stopping is present and hence the more baryons are produced in the central rapidity region. 
In Table\ \ref{tab:aamrs}, $\langle\Delta y\rangle$ and $\langle x \rangle$ and the integrated yield $\langle N_{p} \rangle = \int dy (dN/dy)$ calculated by our model are shown in comparison with experimental values.
The rapidity shifts in $S$ + $S$ reactions for various triggers are in good agreement with the experimental values.
The $\langle\Delta y\rangle$ in both central triggers are significantly larger than one in peripheral trigger, which implies the larger stopping in central collisions than in peripheral collisions.
The rapidity shifts $\langle \Delta y \rangle$ in $Au$ + $Au$ collision at $\sqrt s = 200$ AGeV is much larger than that in $Pb$ + $Pb$ collision at 17.4 AGeV, although $\langle m \rangle$ in two processes are similar to each other.
This increase of $\langle \Delta y \rangle$ is due to the increase of  $\langle m_{3} \rangle$.
In spite of the increase of $\langle\Delta y\rangle$ at RHIC energy region, the baryon number density decreases at the central rapidity region because the beam rapidity also increases.
The mean fractional momentum $\langle x \rangle$ in central collisions of heavy nuclei is smaller than that in central collisions of light nuclei by about 0.1.
However, still more than 20 \% of the incident momentum is carried by the leading nucleons in central collisions of heavy nuclei.
\par

The mean proton multiplicity $\langle N_{p} \rangle$ calculated from our model for $S$ + $S$ collision is in good agreement with data within experimental errors.
It is remarkable that $\langle N_{p} \rangle$ in central collisions of heavy nuclei is larger than the number of incident protons.
This is due to the charge asymmetry of incident nuclei. Consider, for example, collisions of fictitious nuclei made of only neutrons.

\newpage

\section{Conclusions and Discussions}
\label{sec:cd}
It has been shown that our theoretical results of proton spectra are in good agreement with experimental data for both the fractional momentum distribution in $pA$ reactions and the rapidity distributions in central $^{32}$S + $^{32}$S collision at $E_{\mbox{lab}}= 200 AGeV$ and central $^{208}\mbox{Pb}+^{208}\mbox{Pb}$ collision at $E_{\mbox{lab}}=160 AGeV$ (CERN-SPS).
This result suggests strongly the validity of the constituent quark model.
It is predicted that the central rapidity region in $^{197}$Au+$^{197}$Au collision at RHIC energy region will be nearly baryon-free.
In general the baryon number density in the central rapidity region increases with increasing mass number of colliding nuclei, whereas it decreases with increasing incident energy.
During the course of this analysis, we have established a formula, Eq.\ (\ref{eqn:e7}), for the probability of having three types of wounded nucleons in $AB$ collisions and have determined the fragmentation functions, Eq.\ (\ref{eqn:allx11x13}), for those wounded nucleons.
The probabilities given by Eqs.\ (\ref{eqn:d10}), (\ref{eqn:f5}) and (\ref{eqn:e7}), of having different quark interactions in $NN$, $NA$ and $AB$ collisions should be useful to evaluate  various spectra, e.g., the transverse energy distribution, in nuclear collisions at quark level.
\par

It is worthwhile to compare our model with other models for nuclear stopping.
For S $+$ S reaction, the experimental data show a flat rapidity distribution of protons in agreement with our model and VENUS \cite{wer93} while HIJING \cite{xnw91} shows stronger transparency.
For  Pb $+$ Pb collision, our model and RQMD \cite{sor89} predict similar strong stopping \cite{esu95} contrary to VENUS and HIJING which give weaker stopping.
The different models lead to the different predictions for the nuclear stopping.
It should be stressed that the proton spectra of $AB$ collisions in our model result from fitting the $pp$ and $pA$ data through Eq.\ (\ref{eqn:allx2x4}).
In general, any reasonable extrapolation from $pp$ to $AB$ via $pA$ will give a similar result, a rather strong nuclear stopping.
\par
 
Our model can describe the proton spectra in pp, pA and AB collisions in a unified manner.
The notable feature of the fundamental formula, Eq.\ (\ref{eqn:allx2x4}), not shared with other models, is the factorization of x- and A-dependences.
Equation (\ref{eqn:allx2x4}) summarizes the crucial feature of our dynamical assumptions that different types of wounded nucleons distinguished by the number of interacted quarks hadronize differently and independently.
One has to notice that fragmentation functions which have been determined by the experimental data in this work should be eventually derived from theoretical considerations in future.
Anyway, we would like to stress that the constituent quark model is so simple that it has only a few number of free adjustable parameters or functions; $\sigma_{inel}^{qq}$ and $f_{i}(x)$ for $i= 1$, 2 and 3. 
Although most models have recently been constructed as complicated event generators, we feel it still worthwhile to pursue a simple phenomenological model which allows an analytical calculation and a simple interpretation of a result of data analysis.

\section*{Acknowledgment}
The authors would like to thank S. Esumi for kind correspondence on NA44 data of lead-lead collisions(CERN).
One of the authors(T.K.C.) is grateful to Japanese Government for the Monbusho Scholarship. 

\newpage


\appendix
\section{The probabilities of having \lowercase{j} interacted quarks in \lowercase{n} nucleon-nucleon collisions}
\label{sec:appa}
In $NA$ collisions, we have considered that the incident nucleon collides with many nucleons of the target nucleus by multiple collision. 
By expanding the formula of total inelastic cross section Eq.\ (\ref{eqn:f2}), the probabilities of having $n$ (nucleon-nucleon) collisions in an average collision with all possible impact parameters is written as
\begin{equation}
P_{NA}(n)=
\frac{1}{\sigma^{NA}_{inel}}
\int d\bbox{b}
{{A}\choose{n}}
\left[V_{A}(\bbox{b})\right]^{n}
\left[1-V_{A}(\bbox{b}) \right]^{A-n}.
\label{eqn:ap1}
\end{equation}
To obtain the probabilities of projectile nucleon having $j$ interacted quarks in $n$ collisions, it is enough to expand only the second factor of Eq.\ (\ref{eqn:ap1}) $\left[V_{A}(\bbox{b}) \right]^{n}$, the probability of having exactly $n$ collision, in terms of $V_{A}^{(i)}(\bbox{b})$, the probability of finding $i$ interacted quarks in one nucleon-nucleon collision, as follows;
\begin{equation}
\left[V_{A}(\bbox{b})\right]^{n} 
= \sum_{k=0}^{n}\sum_{l=0}^{n-k}{n\choose k}{{n-k}\choose l}\{V_{A}^{(1)}(\bbox{b})\}^{k}\{V_{A}^{(2)}(\bbox{b})\}^{l}\{V_{A}^{(3)}(\bbox{b})\}^{n-k-l}.
\label{eqn:ap2}
\end{equation}   
\par

The factor $\{V_{A}^{(i)}(\bbox{b})\}^{k}$ on the r.h.s  imply the probability of having $k$ collision with the same $V_{A}^{(i)}(\bbox{b})$ and can be expressed in terms of $U_{(i)}^{(j)}(k;\bbox{b})$, the probabilities of having $j$ interacted quarks in {\it{$k$ collisions}} while having $i$ interacted quarks in {\it{each nucleon-nucleon collision}}.
For $k$ collisions with the same probability $V_{A}^{(1)}(\bbox{b})$, it is given by 
\begin{equation}
\{V_{A}^{(1)}(\bbox{b})\}^{k}=
U_{(1)}^{(1)}(k;\bbox{b})
+U_{(1)}^{(2)}(k;\bbox{b})
+U_{(1)}^{(3)}(k;\bbox{b}),
\label{eqn:ap3}
\end{equation}
where
\begin{eqnarray}
U_{(1)}^{(1)}(k;\bbox{b})&=&3\left(\frac{1}{3}V_{A}^{(1)}(\bbox{b})\right)^{k}, 
\label{eqn:ap4} \\
U_{(1)}^{(2)}(k;\bbox{b})&=&3(2^{n}-2)\left(\frac{1}{3}V_{A}^{(1)}(\bbox{b})\right)^{k},
\label{eqn:ap5}    \\
U_{(1)}^{(3)}(k;\bbox{b})&=&\{(3^{n}-3)-3(2^{n}-2)\}\left(\frac{1}{3}V_{A}^{(1)}(\bbox{b})\right)^{k}.
\label{eqn:ap6}
\end{eqnarray}
Here, Eq.\ (\ref{eqn:ap4}) is the probability of one quark in the incident nucleon interacting repeatedly with any nucleon of target nucleus and then having one interacted quark in $k$ collisions.
Eq.\ (\ref{eqn:ap5}) implies the probability of one quark interacting with the target nucleon, also another quark interacting in other collision, then having two interacted quarks.
The probability of three quarks interacting in each collision is appeared in Eq.\ (\ref{eqn:ap6}).
When the projectile nucleon interacts with the target nucleon simultaneously in one collision, there are two possibilities of having two interacted quarks and three interacted quarks in $l$ collision as follows;
\begin{equation}
\{V_{A}^{(2)}(\bbox{b})\}^{l}=
U_{(2)}^{(2)}(l;\bbox{b})
+U_{(2)}^{(3)}(l;\bbox{b}),
\label{eqn:ap7}
\end{equation}
where
\begin{eqnarray}
U_{(2)}^{(2)}(l;\bbox{b})&=&3\left(\frac{1}{3}V_{A}^{(2)}(\bbox{b})\right)^{l}, 
\label{eqn:ap8}  \\
U_{(2)}^{(3)}(l;\bbox{b})&=&(3^{n}-3)\left(\frac{1}{3}V_{A}^{(2)}(\bbox{b})\right)^{l}.
\label{eqn:ap9}
\end{eqnarray}
In case where all the three incident quarks  are participating in one $N-N$ collision, the projectile nucleon has necessarily three interacted quarks in $m$ $N-N$ collisions;
\begin{equation}
\{V_{A}^{(3)}(\bbox{b})\}^{m}=
U_{(3)}^{(3)}(m;\bbox{b}),
\label{eqn:ap91}
\end{equation}
\par

Substituting Eqs.(\ref{eqn:ap3}), (\ref{eqn:ap7}) and (\ref{eqn:ap91}) for Eq.\ (\ref{eqn:ap2}), we obtain
\begin{eqnarray}
\left[V_{A}(\bbox{b})\right]^{n} 
&=&\sum_{k=0}^{n}\sum_{l=0}^{n-k}{n\choose k}{{n-k}\choose l}
\{U_{(1)}^{(1)}(k;\bbox{b})
+U_{(1)}^{(2)}(k;\bbox{b})
+U_{(1)}^{(3)}(k;\bbox{b})\} \nonumber \\
&&\mbox{   }\quad\quad
\times
\{U_{(2)}^{(2)}(l;\bbox{b})
+U_{(2)}^{(3)}(l;\bbox{b})\}
\{U_{(3)}^{(3)}(n-k-l;\bbox{b})\},
\label{eqn:ap10}
\end{eqnarray}   
In Eq.\ (\ref{eqn:ap10}), all terms including $j=3$ contribute the probability having three interacted quarks in $n$ collisions and the cross terms are divided into the probabilities of having two or three interacted quarks with suitable weight.
Thus one can write $\left[V_{A}(\bbox{b})\right]^{n} $  in terms of $U_{NA}^{(j)}(n;\bbox{b})$, the probability of having $j$ interacted quarks in $n$ nucleon-nucleon collision in the following form,
\begin{equation}
\left[V_{A}(\bbox{b})\right]^{n} 
= U_{NA}^{(1)}(n;\bbox{b})+U_{NA}^{(2)}(n;\bbox{b})+U_{NA}^{(3)}(n;\bbox{b}) 
\label{eqn:ap11}
\end{equation}   
where
\begin{eqnarray}
U_{NA}^{(1)}(n;\bbox{b})
&=& U_{(1)}^{(1)}(n;\bbox{b}),\\
U_{NA}^{(2)}(n;\bbox{b})
\label{eqn:ap12}
&=& U_{(1)}^{(2)}(n;\bbox{b})+U_{(2)}^{(2)}(n;\bbox{b}) \nonumber \\
&&\mbox{   }+\sum_{k=1}^{n-1}{n\choose k}
\left[\frac23U_{(1)}^{(1)}(k;\bbox{b})U_{(2)}^{(2)}(n-k;\bbox{b})
+\frac13U_{(1)}^{(2)}(k;\bbox{b})U_{(2)}^{(2)}(n-k;\bbox{b})\right],\\
U_{NA}^{(3)}(n;\bbox{b}) 
\label{eqn:ap13}
&=& \left[V_{A}(\bbox{b})\right]^{n} 
-U_{NA}^{(1)}(n;\bbox{b})-U_{NA}^{(2)}(n;\bbox{b}).
\label{eqn:ap14}
\end{eqnarray}   
From Eqs.(\ref{eqn:ap11}) and (\ref{eqn:ap1}), therefore, we obtain the formulas for probability function of quark absorption in average $n$ collision .

\section{Calculation of the flavor factors in AA collisions}
\label{sec:appb}
The flavor factors in $AA$ collisions is the weighted average of the probabilities that a proton is produced from an incident proton or from an incident neutron.
If one neglects the effect of baryon-antibaryon pair production and assumes that an incident nucleon fragments into either $p$, $n$ or $\Lambda$ one has
\begin{equation}
\lambda_{i}^{(A)} = \frac{Z}{A}(\lambda_{i})+(1-\frac{Z}{A})(1-\lambda_{i}-\eta),
\label{eqn:apb4}
\end{equation}   
where $\eta$ is the probability that an incident nucleon is converted into $\Lambda$ after interaction and $Z/A$ is the proportion of the proton over the atomic mass. 
For a charge symmetric system($Z/A=1/2$), it reduces to
\begin{equation}
\lambda_{1}^{(A)} = \lambda_{2}^{(A)} = \lambda_{3}^{(A)} = \frac{1}{2}(1-\eta).
\label{eqn:apb5}
\end{equation}   
The numerical value of $\eta$ is estimated to be $0.17$ by using the experimental data on central $S + S$ collisions  \cite{bac94,bar90}.

\newpage


\newpage


\begin{figure}
\caption{Three possible interactions of the incident nucleon colliding with the target, nucleon or nucleus, in CQM; one(a), two(b) and all the three(c) incident quarks are interacted with the target.}
\label{fig:figppa}
\end{figure}

\begin{figure}
\caption{The A dependence of the probabilities that one, two or three quarks in an incident nucleon interact in pA collision. }
\label{fig:pai}
\end{figure}

\begin{figure}
\caption{The probability distribution $P_{AB}(m)$ of having $m$ wounded nucleons for different triggers in (a) $^{32}S$ + $^{32}S$ and (b) $^{208}Pb$ + $^{208}Pb$. Here we have used $\sigma_{inel}^{qq} = 4.32$ mb.}
\label{fig:pm}
\end{figure}

\begin{figure}
\caption{The proportion of three types of wounded nucleons $m_{j}$ to total wounded nucleons $m$ in no-bias event of (a) $^{32}S$ + $^{32}S$ and (b) $^{208}Pb$ + $^{208}Pb$ collisions.}
\label{fig:pmi}
\end{figure}

\begin{figure}
\caption{The fragmentation functions $F_{i}(x)$ obtained from the fractional momentum distribution of protons in $pp$, $pCu$ and $pAg$ reactions.}
\label{fig:frg}
\end{figure}

\begin{figure}
\caption{The proton spectra in pA collisions for A = p(a), Be(b), Cu(c), Ag(d), W(e) and U(f); pp, pCu and pAg data are used as inputs. Curves show the model result while data points are taken from \protect\cite{bre82}( $p_{T}$-integrated pp data at 100 GeV/c; triangles), \protect\cite{bai85}( $p_{T}$-integrated pA data at 100 GeV/c; solid circles) and \protect\cite{bar83}( $p_{T}$-fixed pp and pA data at 120 GeV/c; open squares).}
\label{fig:pax}
\end{figure}

\begin{figure}
\caption{The rapidity distribution of participant protons in  $^{32}S + ^{32}S$ collision at $E_{lab} = 200$ AGeV \protect\cite{bac94}. The solid, broken and dotted lines correspond, respectively, to Veto, $E_{T}$ and peripheral triggers. The corresponding experimental data are shown the circles, triangles and squares, respectively.}
\label{fig:ys}
\end{figure}

\begin{figure}
\caption{The proton rapidity distribution in central (15\% trigger) $^{208}Pb + ^{208}Pb$ collision at $E_{lab} = 160$ AGeV . Solid line stands for theoretical result which includes the contribution from $\Lambda$ decay. Experimental data are taken from Ref.\ \protect\cite{esu95}.}
\label{fig:ypb}
\end{figure}

\begin{figure}
\caption{Theoretical prediction for proton rapidity distribution in central (15\% trigger) $^{197}Au + ^{197}Au$ collision at $ \protect\sqrt{s} = 200$ AGeV. Contribution from $\Lambda$ decay is included. It amounts to some 20\% of the total yield.}
\label{fig:yau}
\end{figure}

\newpage


\begin{table}
\caption{Comparison of a calculated total inelastic cross section $\sigma^{pA}_{inel}$ with the experimental values from Ref. \protect\cite{bai85} and \protect\cite{car79}, the probabilities $P_{pA}^{(i)}$ having one, two or all the three interacted quarks in $pA$ collisions and their average value $<i>$.}
\label{tab:pacp}
\begin{center}
 \begin{tabular}{ccccccc} 
 {A} & {${\sigma^{pA}_{inel}}|_{cal}(mb)$} & {${\sigma^{pA}_{inel}}|_{exp}(mb)$ } & {$P_{pA}^{(1)}$} & {$P_{pA}^{(2)}$} & {$P_{pA}^{(3)}$} & {$<i>$} \\ \hline
$p$        & 30     & 31.3$\pm$1.2 & 0.81 & 0.17 & 0.02  &  1.21    \\
$^{9}Be$   & 188    & 176$\pm$2    & 0.61 & 0.30 & 0.09  &  1.48    \\
$^{32}S$   & 493    &    $-$       & 0.47 & 0.34 & 0.19  &  1.72    \\
$^{64}Cu$  & 785    & 767$\pm$8    & 0.39 & 0.34 & 0.27  &  1.88    \\
$^{108}Ag$ & 1118   & 1097$\pm$12  & 0.33 & 0.33 & 0.34  &  2.01    \\
$^{189}W$  & 1584   & 1540$\pm$16  & 0.27 & 0.31 & 0.42  &  2.15    \\
$^{208}Pb$ & 1724   & 1752$\pm$53  & 0.26 & 0.30 & 0.44  &  2.18    \\
$^{238}U$  & 1880   & 1860$\pm$20  & 0.25 & 0.29 & 0.46  &  2.21    \\  
 \end{tabular}
\end{center}
\end{table}

\begin{table}
\caption{Comparison of a calculated total inelastic cross section $\sigma^{AA}_{inel}$ with the experimental values from Ref. \protect\cite{bac94} and empirical formula, the average number of wounded nucleons $\langle m \rangle $ and the average values for three type of wounded nucleon $\langle m_{1} \rangle $, $\langle m_{2} \rangle $, $\langle m_{3} \rangle $ for different trigger in $AA$ collision. Here ``no-bias'' implies that the impact parameter integration is carried out from zero to infinity.}
\label{tab:aacp}
\begin{center}
 \begin{tabular}{clccrrrr} 
 {$AA(\sqrt{s}(AGeV))$} & {trigger} & {${\sigma^{AA}_{inel}}|_{cal}(mb)$} & {${\sigma^{AA}_{inel}}|_{exp}(mb)$ } & {$\langle m \rangle $} & {$\langle m_{1} \rangle$} & {$\langle m_{2} \rangle $} & {$\langle m_{3} \rangle$ }\\ \hline
 S+S(19.4)  &no-bias    & 2173 & 1700($1740^{*}$) & 9.4   & 5.6  & 2.9   & 0.9   \\
      &Veto(2\%)    & 43   &    34            & 27.2  & 9.8 & 11.1  & 6.3   \\
      &$E_{T}$(11\%)& 239  &   190            & 25.3  & 10.6 & 9.9   & 4.8   \\
      &periphe.     & 1422  &  1000           & 4.0  & 3.1 & 0.8   & 0.1   \\
 Pb+Pb(17.4)&no-bias    & 7660 &  $7630^{*}$     & 77.0  & 32.7 & 26.6  & 17.8  \\ 
      &cent.(15\%)  & 1149 &   $-$           & 196.0 & 39.5 & 75.5  & 81.0  \\
 Au+Au(200)&no-bias    & 7414($7179^{\dag}$) &  $7325^{*}$     & 80.2  & 25.8 & 25.8  & 28.6  \\ 
      &cent.(15\%)  & 1112 &   $-$           & 192.7 & 18.2 & 54.0  & 120.5 \\ 
 \end{tabular}
\end{center}
{\small $^{*}$Values of empirical formula in Ref. \cite{abd80}, see text.}
{\small $^{\dag}$ Value for $\sigma_{inel}^{qq} = 4.32$ mb.}
\end{table}

\begin{table}
\caption{Flavor factors $\lambda_{i}^{(A)}$ calculated by Eqs.\ (\protect\ref{eqn:apb4}) and (\protect\ref{eqn:apb5}) in Appendix \protect\ref{sec:appb}}
\label{tab:flaf}
\begin{center}
 \begin{tabular}{lccc} 
 {A} & {$\lambda_{1}^{(A)}$} & {$\lambda_{2}^{(A)}$} & {$\lambda_{3}^{(A)}$} 
\\ \hline
$S$    & 0.415  & 0.415  & 0.415    \\
$Pb$   & 0.365  & 0.374  & 0.397    \\
$Au$   & 0.368  & 0.376  & 0.398    \\
 \end{tabular}
\end{center}
\end{table}

\begin{table}
\caption{The mean multiplicities $\langle N_{p}\rangle$ and the mean rapidity shift $\langle \Delta y \rangle$ $$ for $y < y_{cm}$ of the final state proton in our theoretical results are compared with the experimental values in Ref. \protect\cite{bac94} for $S + S$ reaction at $E_{lab} = 200$ AGeV. $\langle x \rangle$ is the mean fractional momentum retained by the final state proton.}
\label{tab:aamrs}
\begin{center}
 \begin{tabular}{clcrcrrccc} 
 {$AA(\sqrt{s}(AGeV))$} & {trigger} & {$ \langle N_{p} \rangle|_{exp}$} & {$\langle N_{p} \rangle|_{cal} $} & {$\langle \Delta y_{p} \rangle |_{exp}$} & {$\langle \Delta y \rangle |_{cal}$} & {$<x>|_{cal}$} \\ \hline
 S+S(19.4) & Veto(2\%) & 12.8$\pm$1.4  & 14.4 & 1.58$\pm$0.15 & 1.52 & 0.34 \\
       & $E_{T}$(11\%) & 10.3$\pm$1.4  & 13.3 & 1.58$\pm$0.15 & 1.46 & 0.36 \\
       & periphe.       & 3.1$\pm$0.8   & 2.0  & 1.00$\pm$0.15 & 1.16 & 0.47 \\
 Pb+Pb(17.4)& cent.(15\%)& $-$         & 96.4 &   $-$         & 1.67 & 0.27 \\
 Au+Au(200) & cent.(15\%) &    $-$     & 96.6 &   $-$         & 2.22 & 0.22 \\
 \end{tabular}
\end{center}
\end{table}

\newpage


\begin{figure}[t]

\centering{\epsfxsize 16cm \epsffile{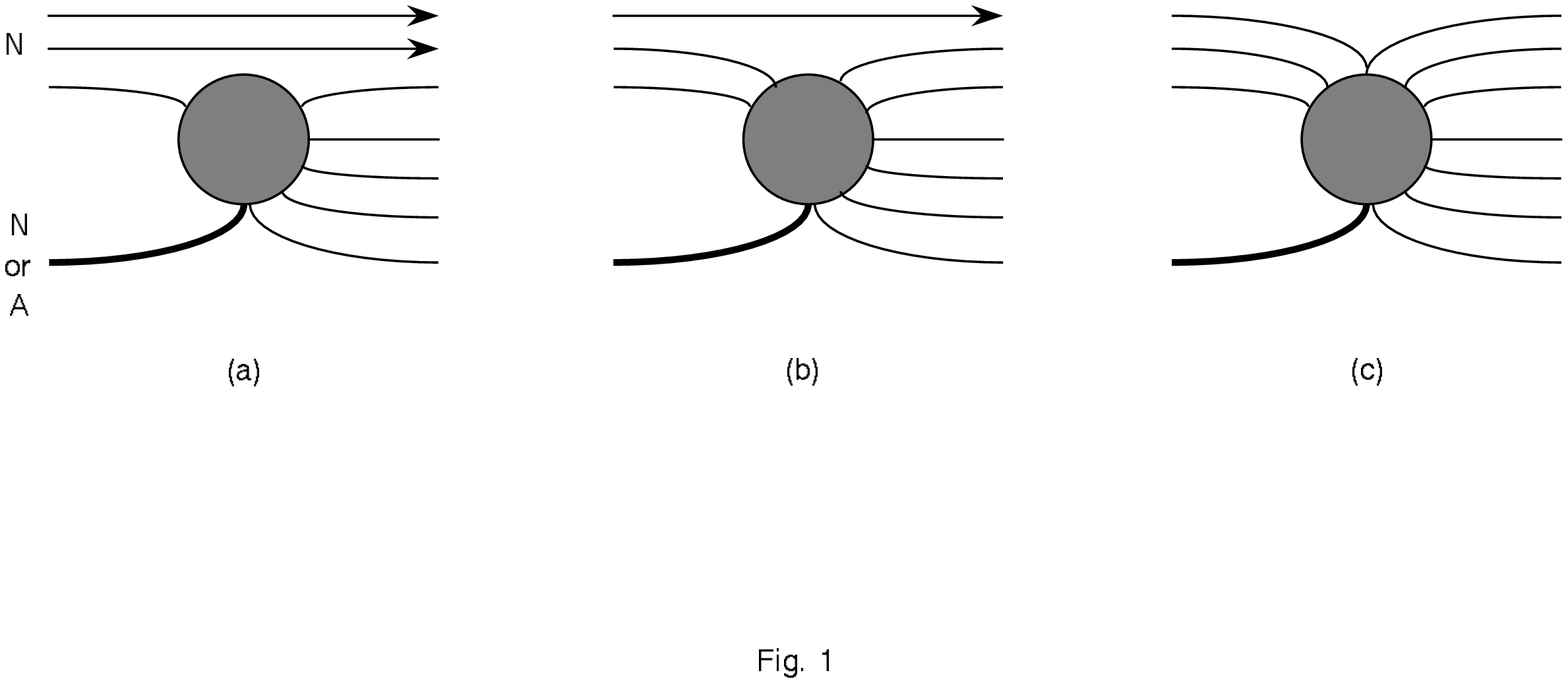}}

\end{figure}

\begin{figure}[b]
\centering{\epsfxsize 10cm \epsffile{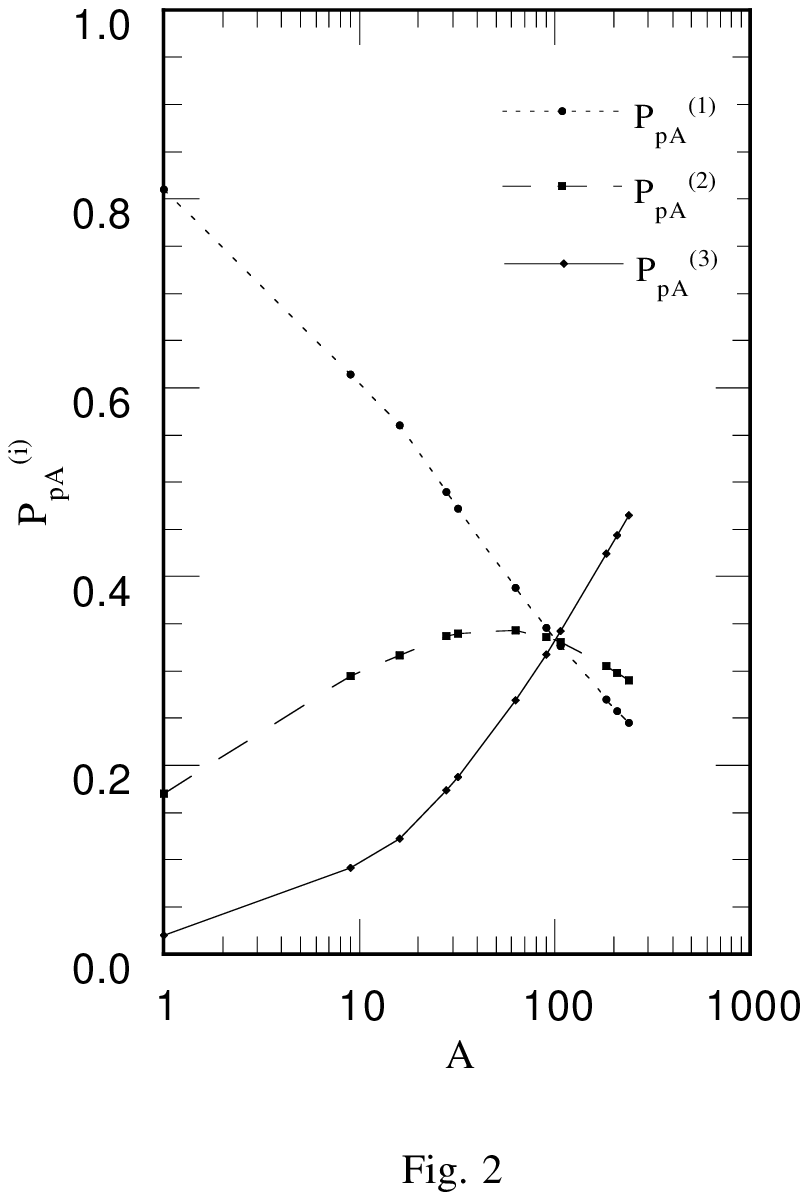}}
\caption{Fig.2}
\end{figure}

\newpage
\begin{figure}[t]
\centering{\epsfxsize 13cm \epsffile{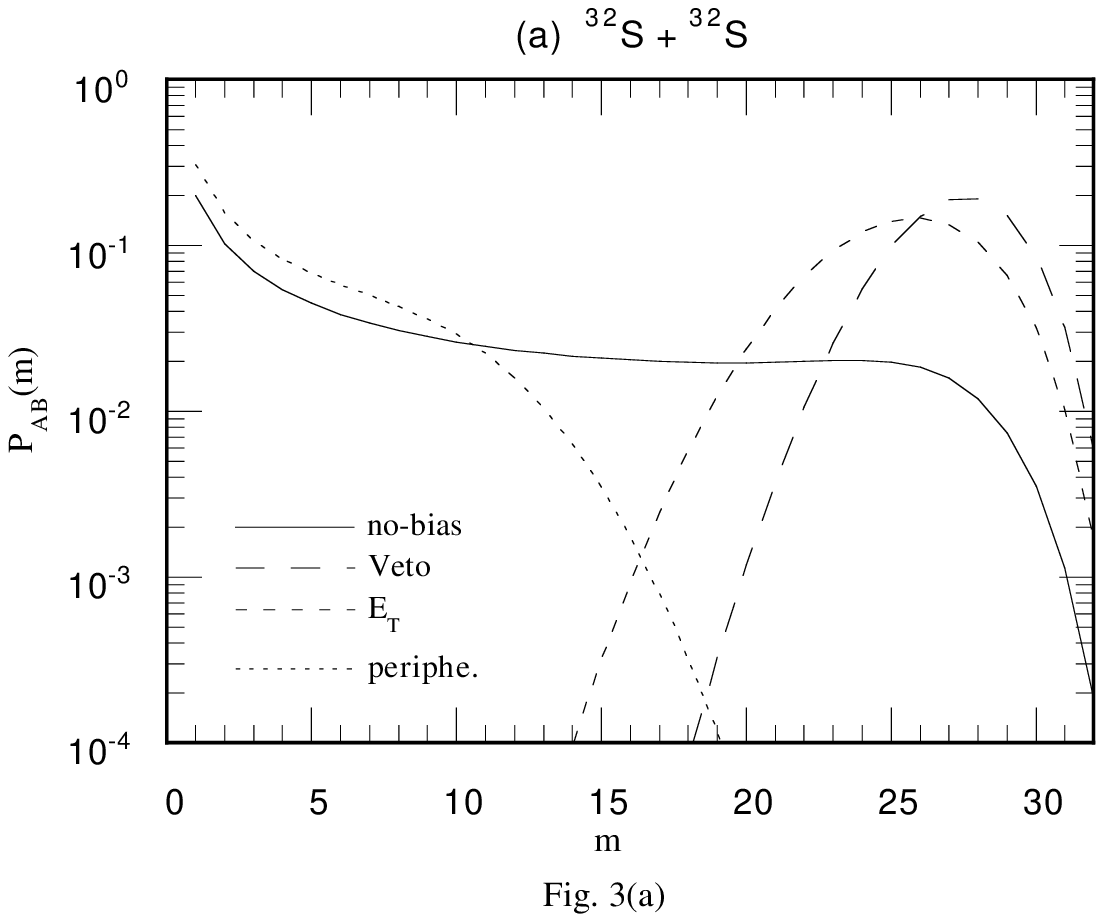}}
\end{figure}

\begin{figure}[b]
\centering{\epsfxsize 13cm \epsffile{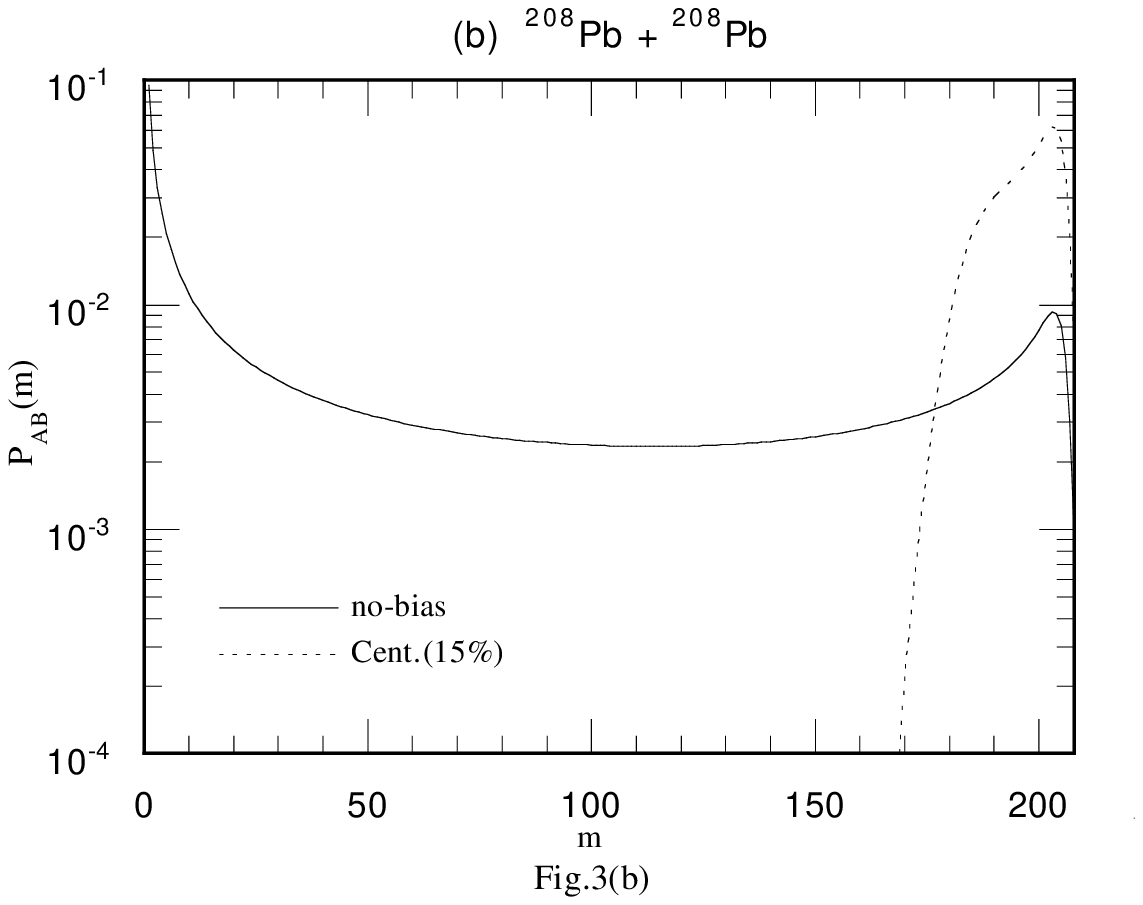}}
\end{figure}

\newpage
\begin{figure}[t]
\centering{\epsfxsize 13cm \epsffile{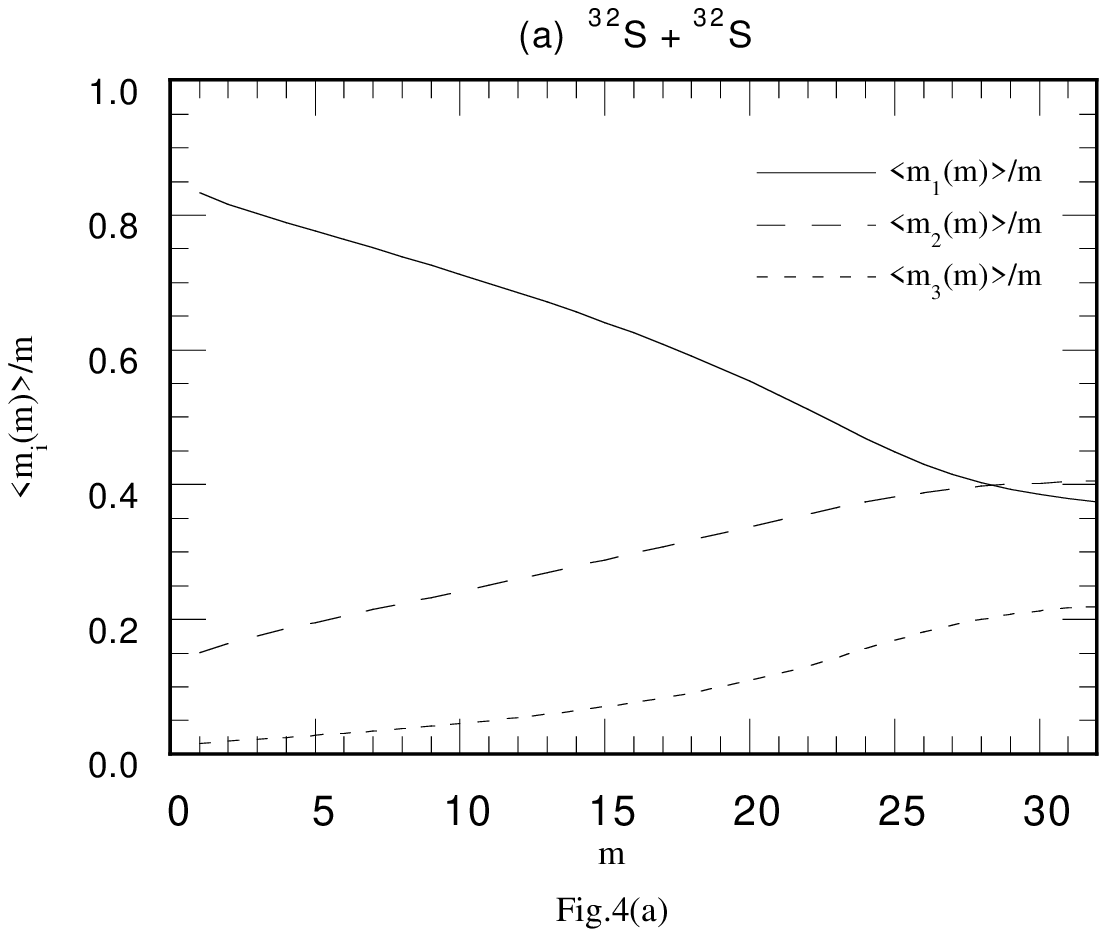}}
\end{figure}

\begin{figure}[b]
\centering{\epsfxsize 13cm \epsffile{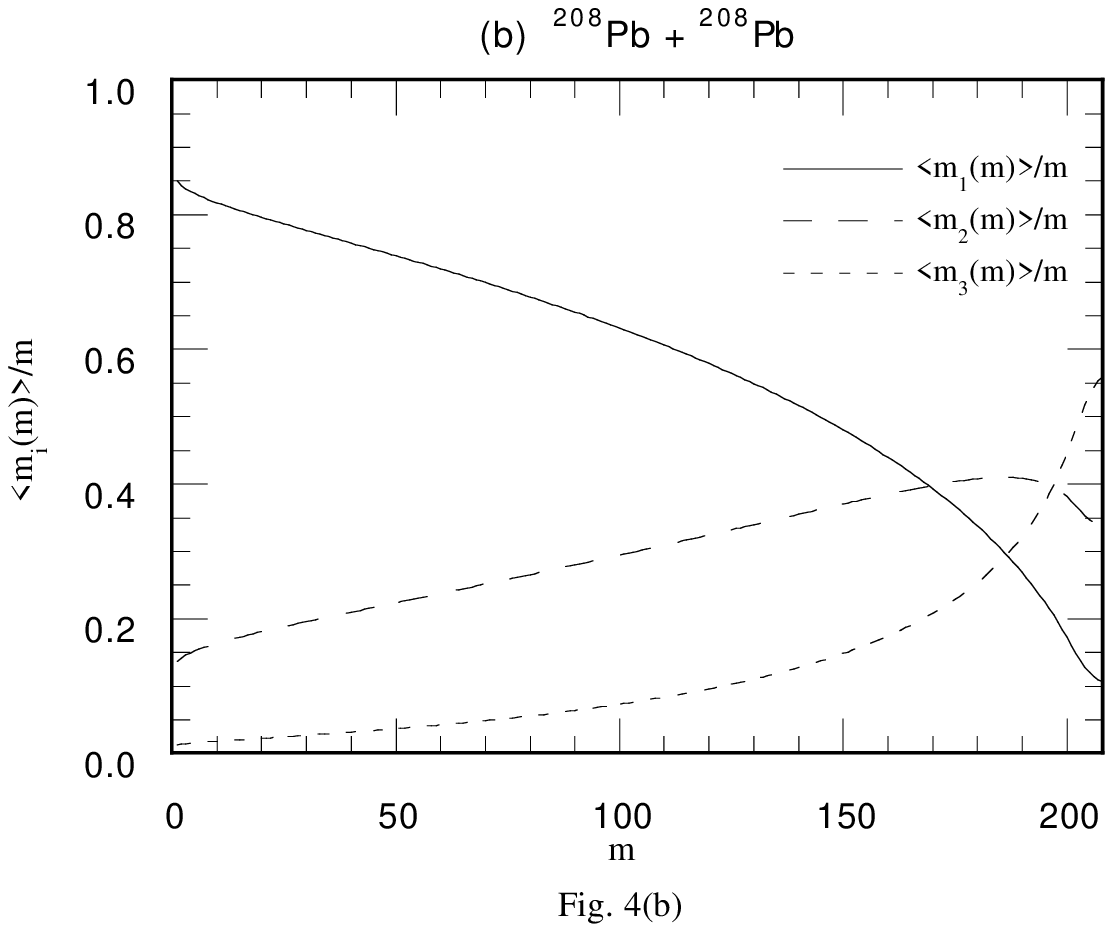}}
\end{figure}

\newpage
\begin{figure}[t]
\centering{\epsfxsize 14cm \epsffile{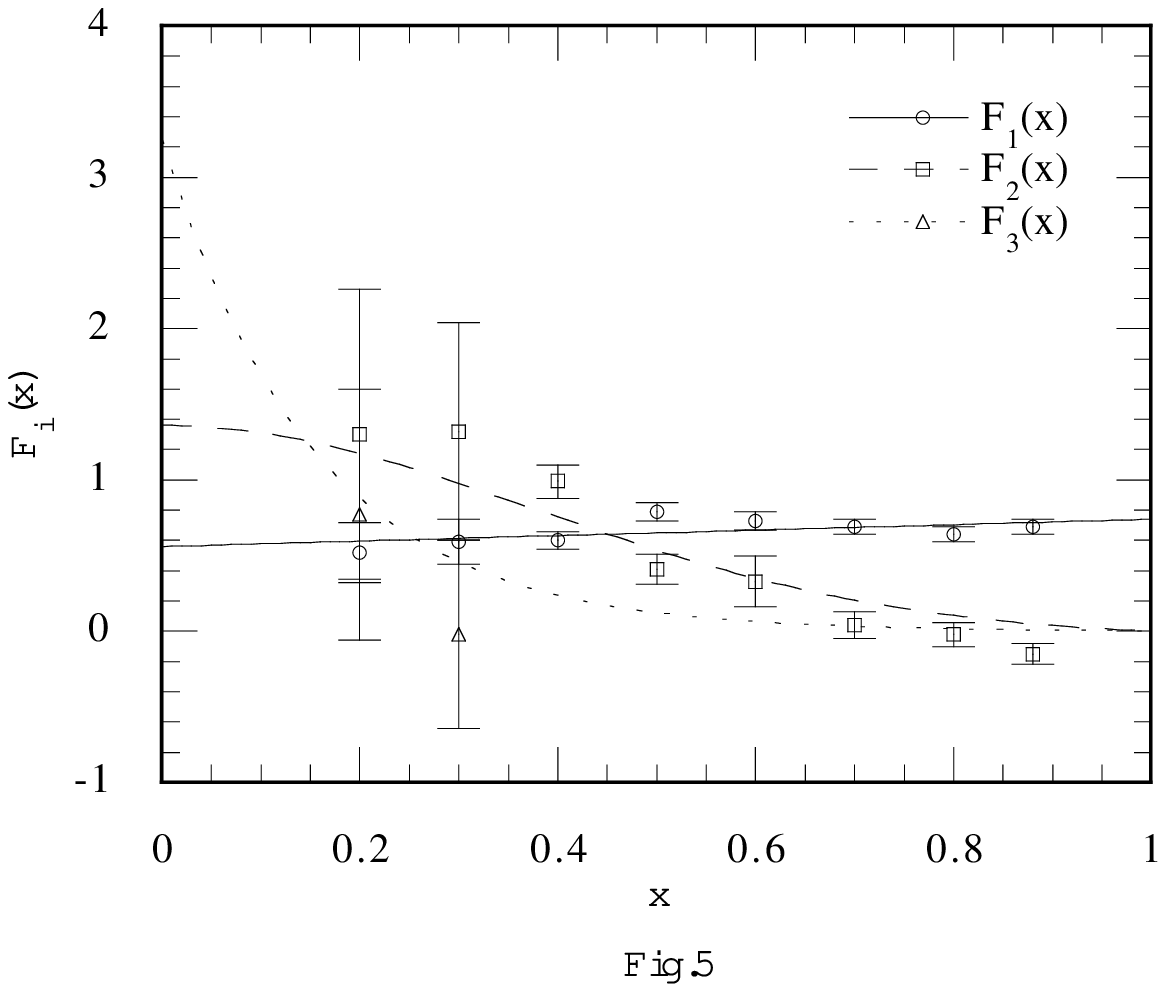}}
\end{figure}

\newpage
\begin{figure}[t]
\centering{\epsfxsize 13.4cm \epsffile{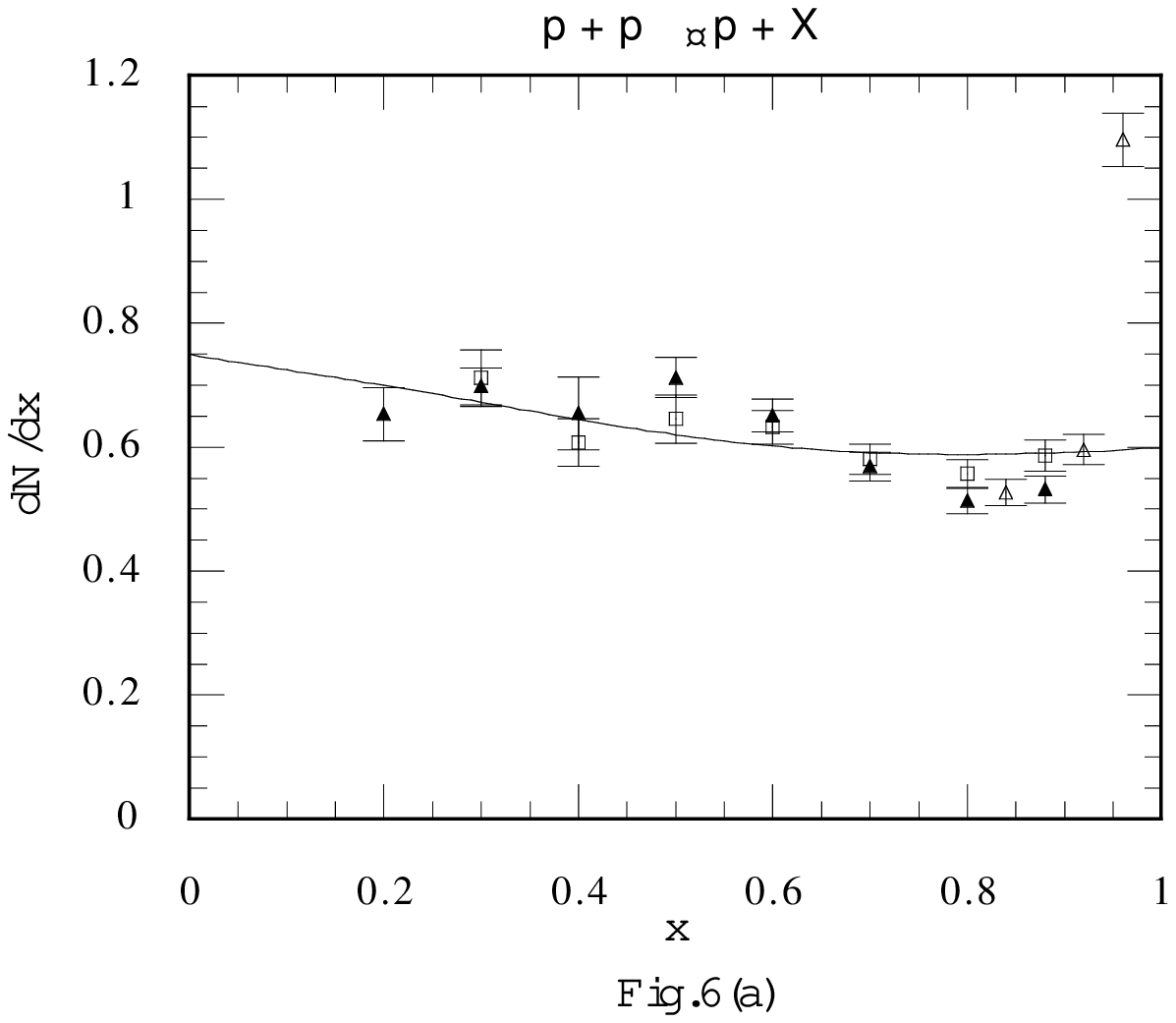}}
\end{figure}

\begin{figure}[b]
\centering{\epsfxsize 14cm \epsffile{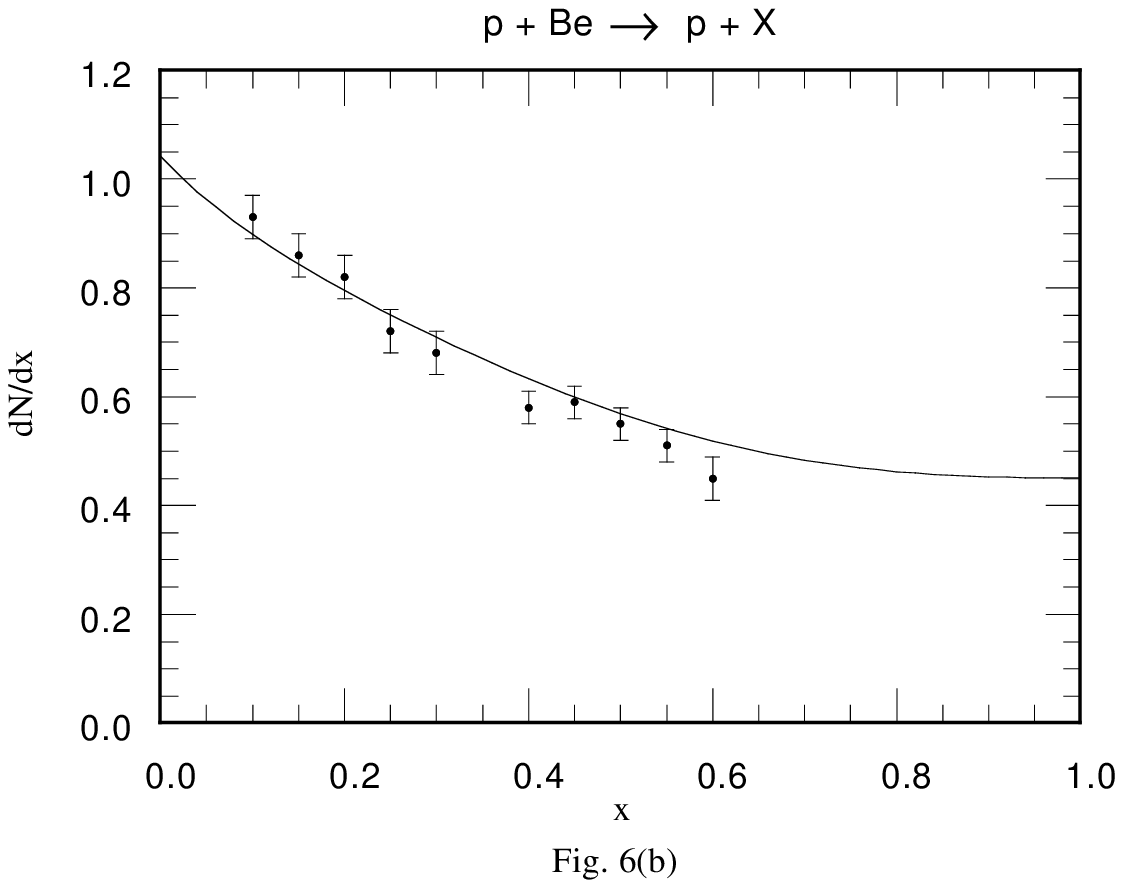}}
\end{figure}

\newpage
\begin{figure}[t]
\centering{\epsfxsize 14cm \epsffile{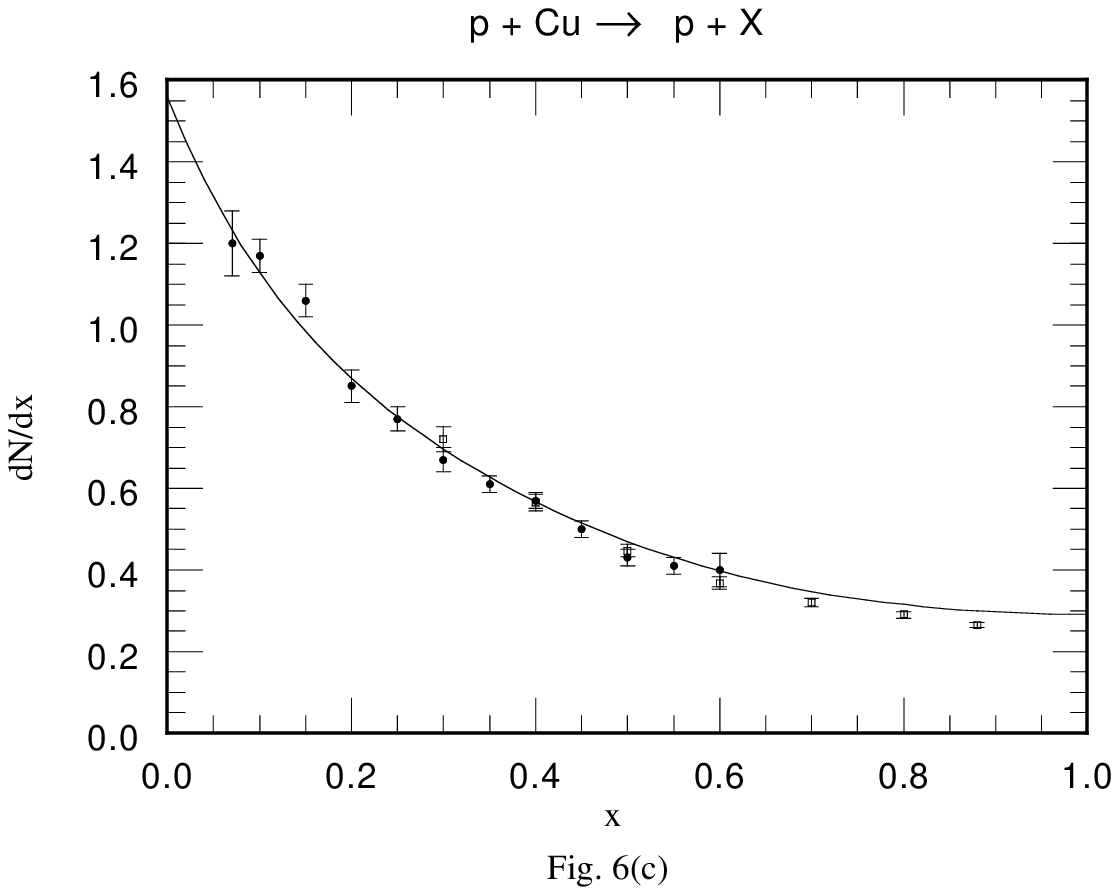}}
\end{figure}

\begin{figure}[b]
\centering{\epsfxsize 14cm \epsffile{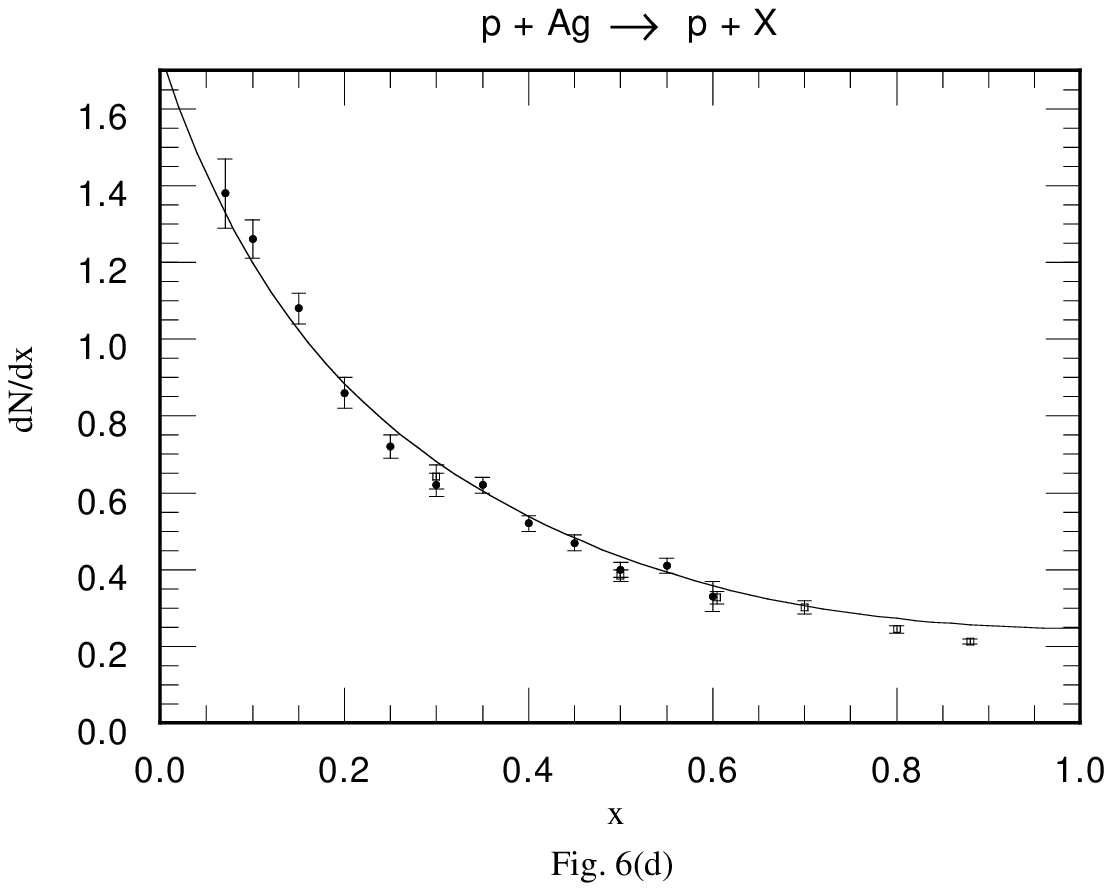}}
\end{figure}

\newpage
\begin{figure}[t]
\centering{\epsfxsize 14cm \epsffile{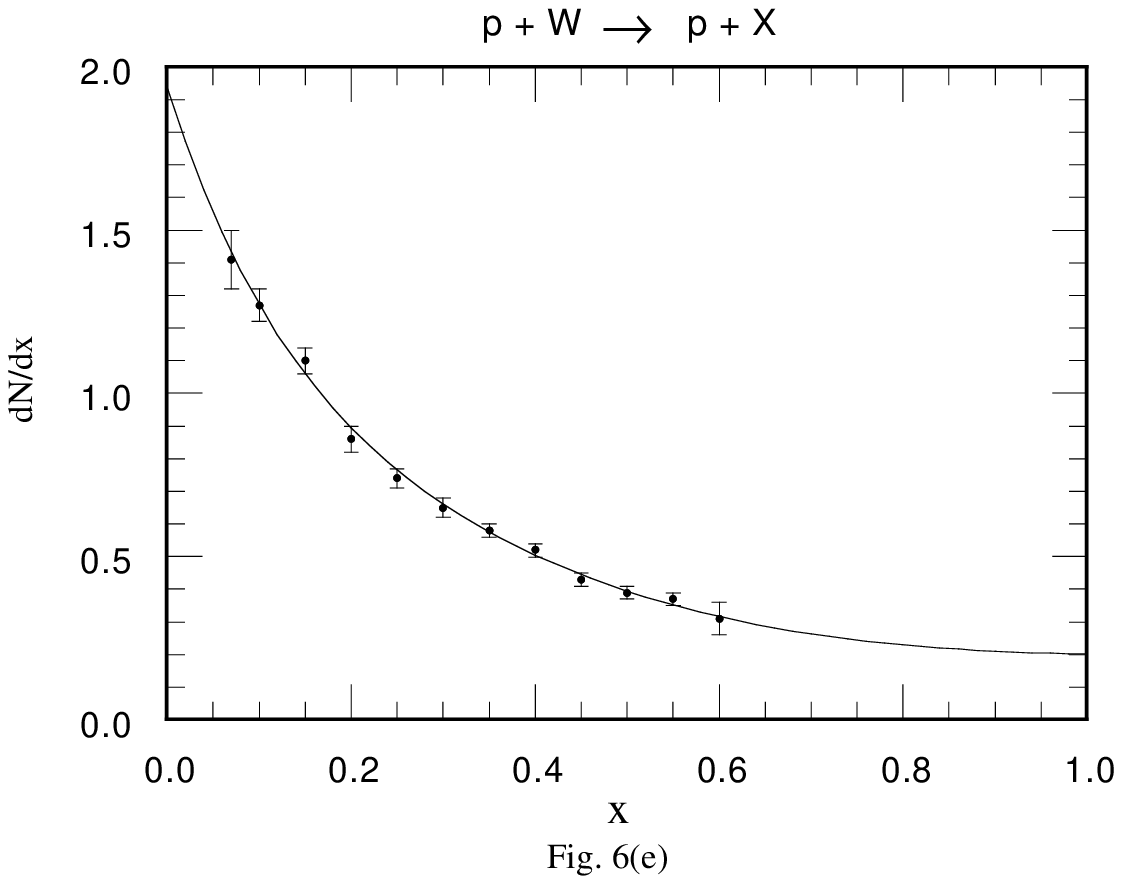}}
\end{figure}

\begin{figure}[b]
\centering{\epsfxsize 14cm \epsffile{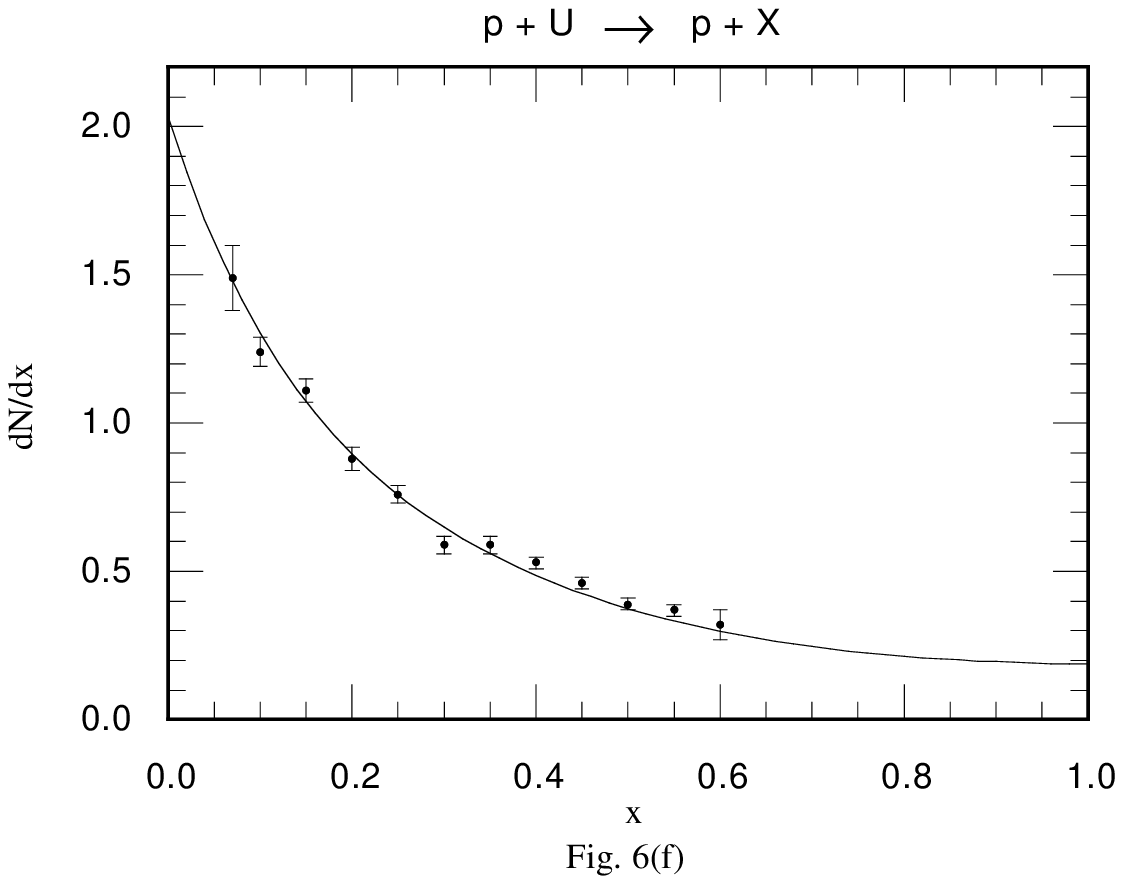}}
\end{figure}

\newpage
\begin{figure}[t]
\centering{\epsfxsize 14cm \epsffile{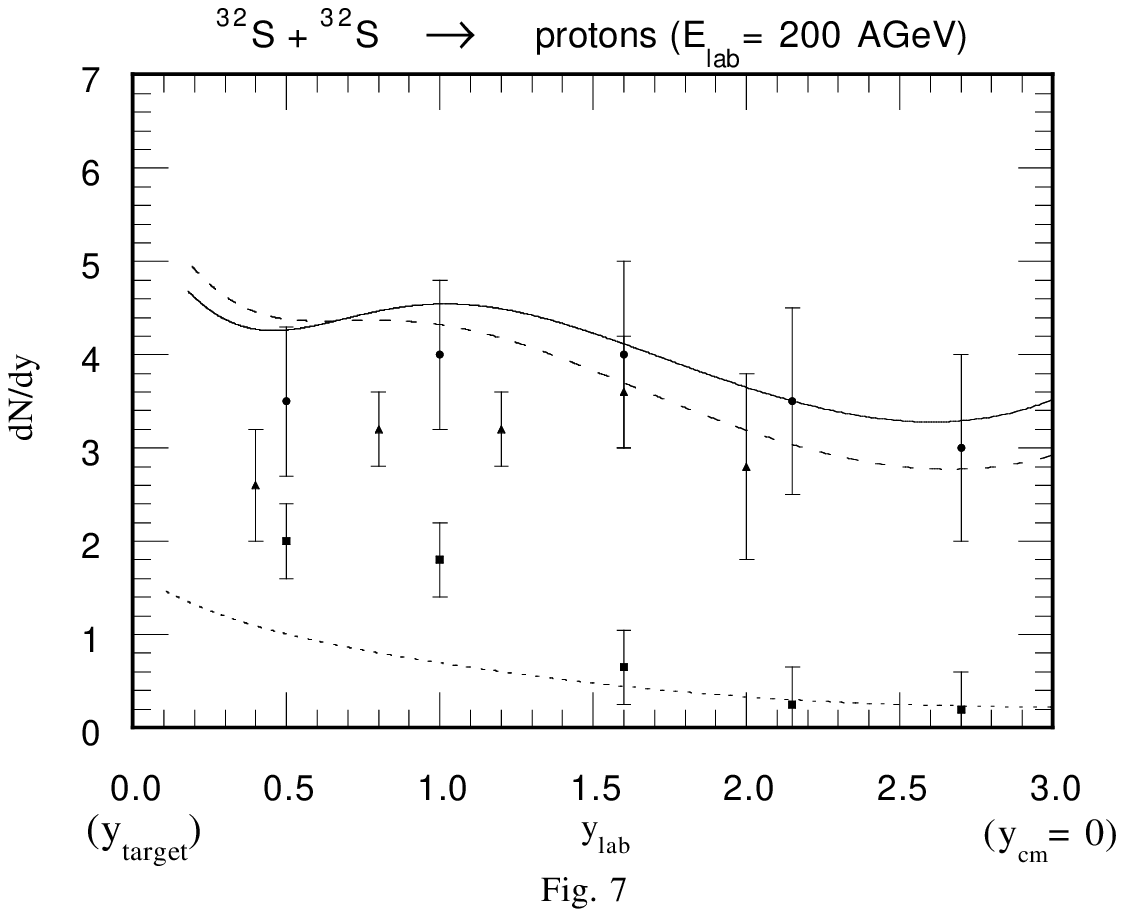}}
\end{figure}

\begin{figure}[b]
\centering{\epsfxsize 14cm \epsffile{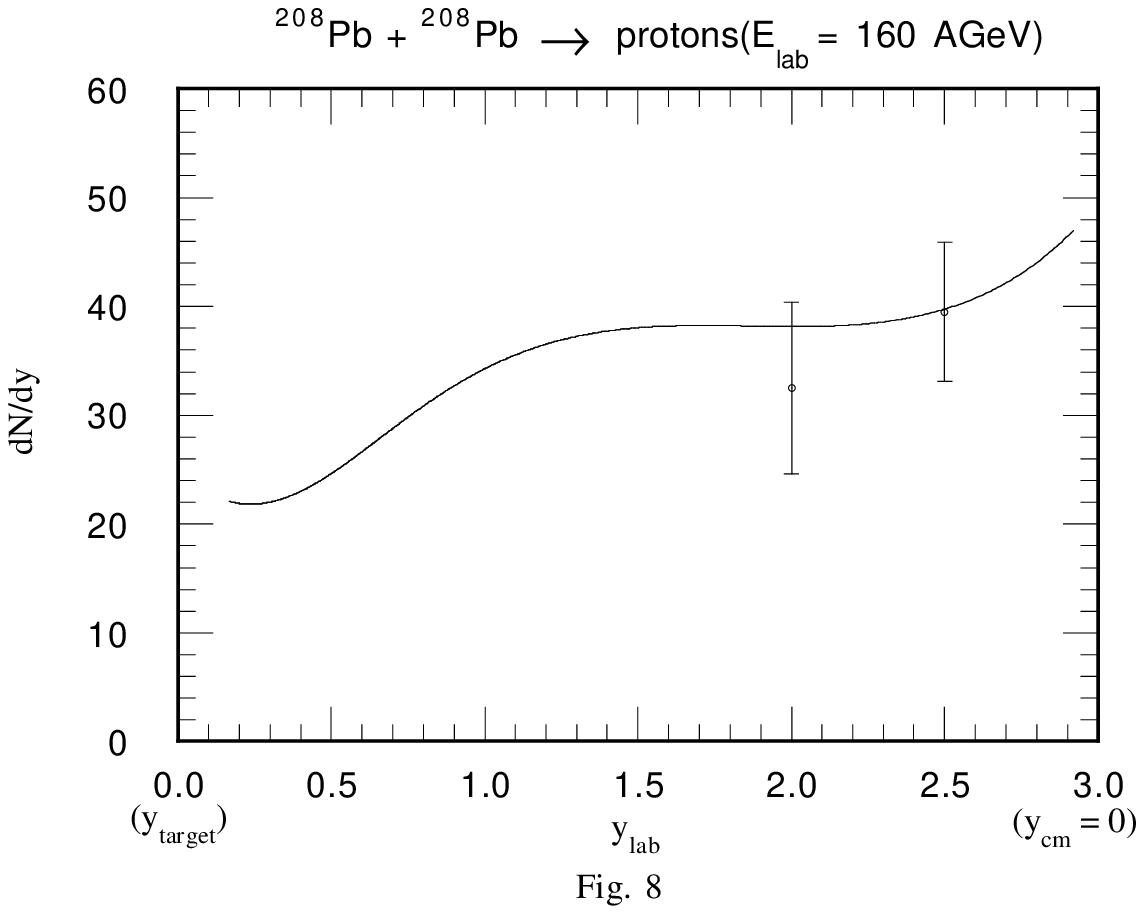}}
\end{figure}

\newpage
\begin{figure}[t]
\centering{\epsfxsize 14cm \epsffile{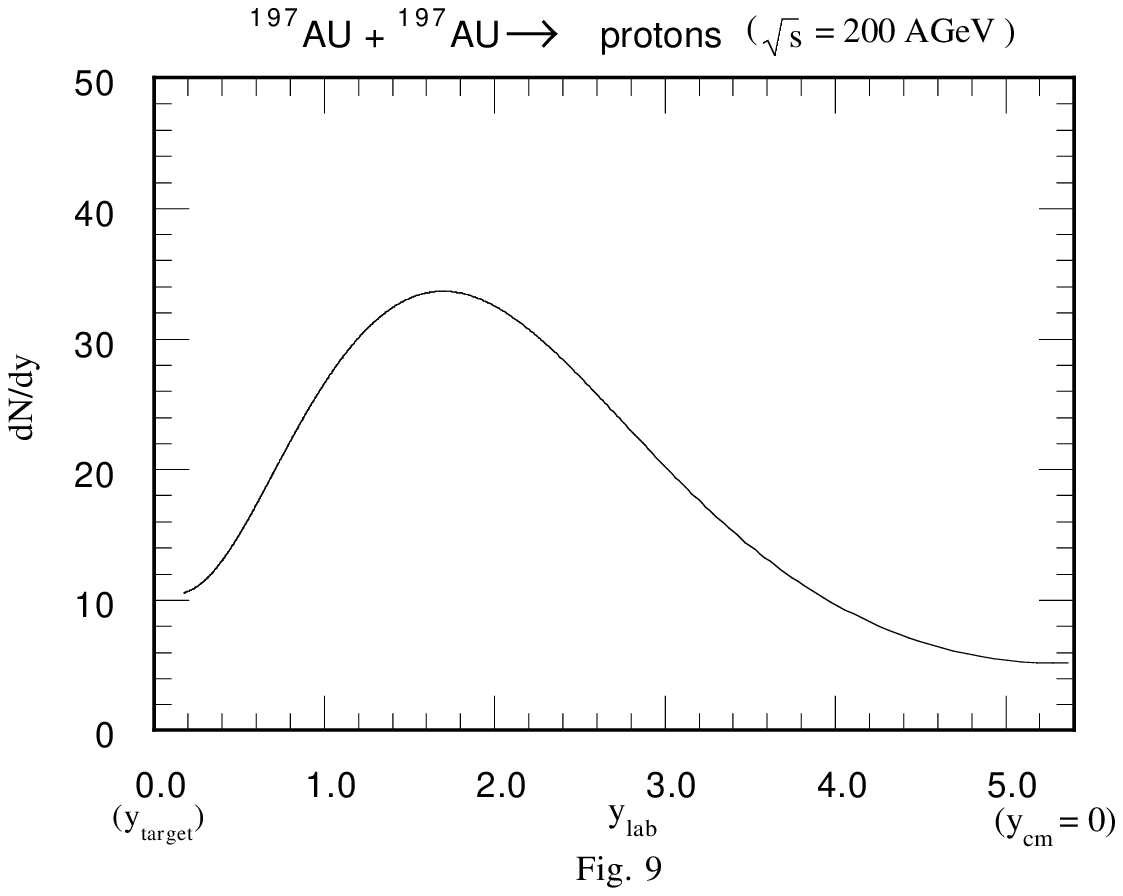}}
\end{figure}

\end{document}